\documentclass{article}
\pdfoutput=1
\usepackage{times}
\usepackage[pdftex]{graphicx}
\begin{document}
\title{Formation of accretion centers in simulations of colliding uniform 
density H$_2$ cores}

\author{G. Arreaga-Garc\'{\i}a\footnote{Corresponding 
author:garreaga@cifus.uson.mx}\\
Departamento de Investigaci\'on en F\'{\i}sica de la Universidad de Sonora \\
Apdo. Postal 14740, Hermosillo, 83000 Sonora, M\'exico.\\
J. Klapp \\
Departamento de F\'{\i}sica, Instituto Nacional de Investigaciones Nucleares,\\
Carretera M\'exico-Toluca, Ocoyoacac, 52750 Estado de M\'exico, M\'exico.}

\maketitle

\abstract{We test here the first stage of a route of modifications  
to be applied to the public GADGET2 code for dynamically identifying
accretion centers during the collision process of two adjacent and
identical gas cores. Each colliding core has a uniform density
profile and rigid body rotation; its mass and size have been chosen
to represent the observed core $L1544$; for the thermal and rotational
energy ratios with respect to the potential energy, we assume the
values $\alpha=0.3$ and $\beta=0.1$, respectively. These values
favor the gravitational collapse of the core. We here study
cases of both -head-on and off-center collisions, in which the
pre-collision velocity increases the initial sound speed of
the barotropic gas by up to several times. In a simulation the
accretion centers are formed by the highest density particles, so we
here report their location and properties in order to realize the
collision effects on the collapsing and colliding cores. In one of
the models we observe a roughly spherical distribution of accretion
centers located at the front wave of the collision. In a forthcoming 
publication we will apply the full modified GADGET code to study 
the collision of turbulent cores.}




\section{Introduction}
\label{intro}

Stars form out of collapsing prestellar cores, located usually in 
larger clumps that could be isolated or part of a larger gas 
filament.  Massive stars could also form as a  result of the collapse
of a relatively 
large core, \cite{zinne}. Besides, the massive star formation 
scenario by means of gas collisions
has gained observational support from recent space observation
projects like ALMA, NANTEN, and the Spitzer telescope, all of which
provide evidence of the occurrence of collisions among gas clouds
and/or clumps, see section 1.1 of \cite{takahira} and references 
there in. 

Besides, \cite{testi} have presented high-resolution
interferometric and single-dish observations of molecular gas in the
Serpens cluster-forming core. It appears that the Serpens core
encompasses at least three substructures, the northwest and
southeast sub-clusters and the near-IR cluster. Thus, the spatially
distinct structures are also kinematically separated, while there is
reasonable internal velocity coherence. Star formation is currently
occurring simultaneously in the northwest and southeast
sub-clusters, both of which contain roughly equal fractions of
pre-stellar, proto-stellar, and IR sources. These facts point to the
idea that Serpens was formed by means of a collision between two or
three very similar gas clumps. \cite{graves} and \cite{duarte} have 
recentely observed the Serpens molecular 
cloud by using HARP at the James Clerk Maxwell Telescope and reported 
evidence that Serperns consists of two colliding sub clouds. 

Furthermore, \cite{furu}, \cite{torii} and \cite{churchwell}, have reported 
about more 
recent observational evidence of cluster formation triggered by cloud-cloud 
collisions. They included the cases of RCW49, Westerlund2 and NGC3603 as 
examples.     
We then assume that collisions between smaller gas 
structures are also possible. 

Turbulence makes a big change in the spatial distribution of the 
interstellar gas as turbulent clouds become filamentary and
flocculent and their gas is locally compressed such that dense cores can form 
and eventually become gravitationally unstable to form stars, see for 
instance \cite{fede2012} and \cite{padoan}. 
It is within highly turbulent molecular
clouds where core collisions may occur because of the random nature of the 
gas velocity favors
the formation of small over-densities across the cloud, which might
be already collapsing or at least about to start collapsing. 

Besides, \cite{motte} have shown evidence that several parts of
the Oph cloud have been compressed by an external pressure, such as a 
supernova explosion,
an expanding HII region, or stellar wind shells. \cite{scoville}
have also presented observational evidence of gas compression at the
interface of two colliding molecular gas regions.
Thus, there is sufficient observational evidence to support the idea that
gas collisions in all their possible size structures can occur. One can then
ask what could be the influence of core collision on star formation.

Collision between relatively smaller clumps of gas and larger clouds
have been studied numerically for more than three  decades, for instance,
\cite{hausman} and \cite{lattanzio}; particularly, head-on
collisions were considered by \cite{kimura}, \cite{klein} and
\cite{marinho}. Some of the earliest works were performed with only a few thousand
particles and therefore suffered from a lack of resolution. However, in the recent
past, simulations have been developed with much higher spatial resolution,  for
instance \cite{burkert} and \cite{pindika}. The case of colliding
gas structures starting from hydrodynamical equilibrium was considered by
\cite{kitsionas} and \cite{pindika2}; \cite{pindika3} has
considered collisions between dissimilar gas structures. 
~\cite{gom} and ~\cite{evaz} and other authors as well, have studied the 
generation of turbulence and star formation at the shock front of head-on 
collisions.

We here present numerical high resolution three 
dimensional hydrodynamical simulations of two rotating gas 
core collisions, including the
self-gravity of the gas. We calculate all collision models with a
modified GADGET2 code (which is describe in Section
~\ref{subs:code}) in order to describe the formation of accretion
centers in a collision system. We must clarify that accretion centers are 
conceptually simlar but not identical to the {\it sink particles}.

The sink particle technique, as implemented by
\cite{bate} and \cite{fede}, has been very useful, as one of
the most important issues in star formation
numerical simulations is that of identifying peak density gas
parcels and monitoring their evolution as closely as possible. In the
case of the particle-based codes used to follow the
gravitational collapse of cores, all the higher density particles
are very closely packed, so that this set of particles is likely
to be identified with a protostar. The sink particles play the role of
proto-stars in the simulations and their physical properties are
easily investigated in this scheme, which hopefully could be
compared with observations. For instance, \cite{clark} have used the 
sink particle technique in their study of the mass function of cores by 
means of colliding clumpy flows.

We devote this paper to making new numerical
simulations of the collision process of two rotating cores, taking advantage 
of both the new capabilities of modern computers and improved numerical techniques.
We emphasize that a core very similar to the one which we take here,
was originally suggested by \cite{whithworth} as an empirical model
to study the collapse of the well-observed core L1544. The physical
properties of this core fit more or less the observed average
properties of the cores reported by ~\cite{tafalla}
and ~\cite{jijina}.  \cite{tafalla} have presented a multi-line and
continuum study of L1544, and according to them this core presents
evidence to be in an advanced state of evolution towards the
formation of one or two low mass stars.

In this paper we take only the core size and mass as given 
by \cite{whithworth}, and we
additionally consider the core to be in counterclockwise rigid
body rotation around the $Z$ axis. We also include a density 
perturbation on the initial core model in order for the isolated 
collapsing core to develop a
filament which eventually would fragment into a binary protostar system.

We here study collision effects on this collapsing core. The 
Rankine-Hugoniot jump condition applied to the case of two colliding cores, 
indicate that the density in the thin dissipative region where 
the collision takes place, must be increased by about four
times the pre-collision density, so we expect an overdensity to be 
formed at the interface region of the colliding
cores and also at the core's central region where gravitational
collapse is ongoing. The nature of the final distribution of
proto-stars will depend on the collision parameters, mainly on
the pre-collision velocity and the impact factor.

A word of warning is in order. In any star formation theory, it is
very difficult to make numerical calculations with realistic initial
conditions. Particularly difficult is the case of core collisions,
where the parameter space is enormous and so we are here forced
(like other authors, as already mentioned in the above paragraphs)
to use simplified initial conditions. Despite this, we assess
what are the effects of a collision on the collapsing and colliding
cores, as well as where and when the accretion centers form, which
are the most important issues to be investigated in this paper.

The outline of this paper is as follows. In Section
\ref{sec:nubeinicial} we describe the initial core, which will be
involved in all the subsequent collision models. In Section
\ref{sec:modelos} we give the details of the collision geometry and
define the models to be studied. We shall also describe the modified
GADGET2 code and the most important free parameters, which we
fix by means of a calibration method presented in Section
~\ref{subs:code}. In Section \ref{sec:resultados}, we describe the
most important features of the time evolution of our simulations by
means of two-dimensional ($2D$) and three-dimensional ($3D$) plots.
In Sections \ref{sec:dis} and \ref{sec:conclu} we discuss the
relevance of our results in view of those reported by previous papers
and finally we make some concluding remarks.

\section{The initial core}
\label{sec:nubeinicial}

We consider a uniform density spherical core model with radius
$R_0=8.0\, \times 10^{16} \,$ cm ($\, =0.026\; $ pc $\, =5350 \; $AU) and 
mass $M_0=8 \; M_\odot$. The average density of this core
is $\rho_0 = 9.2 \, \times 10^{-18}\; $ g $\,$ cm$^{-3} \equiv 2.3
\times 10^{6} \;$ part $\,$ cm$^{-3}$, from which an estimate of its
free fall time is $t_{ff}=\sqrt{3 \pi / (32 G \rho_0) } = 6.9 \times
10^{11} \,$ s $= 21879 \; $yrs.

The dynamical fate of a collapsing isolated core is usually
characterized by the values of the thermal and rotational energy
ratios with respect to the gravitational energy, denoted by $\alpha$
and $\beta$, respectively, see \cite{bodeniv} and 
\cite{sigalotti2001a}, \cite{sigalotti2001b}.
For this paper, all the simulations have been done with an initial core
having the same initial values $\alpha$ and $\beta$

\begin{equation}
\begin{array}{l}
\alpha \equiv \frac{E_{\rm therm}}{\left|E_{\rm grav}\right|} \approx
0.3 \, ,
\vspace{0.25 cm}\\
\beta \equiv \frac{E_{\rm rot}}{\left|E_{\rm grav}\right|}\approx 0.1 \, .
\end{array}
\label{defalphabeta}
\end{equation}

As is well known, these values for $\alpha$ and $\beta$ favor the
occurrence of core collapse. The speed of sound $c_0$ and the
rotational angular velocity $\Omega_0$ have been calculated to
satisfy the energy ratios given by Eq. \ref{defalphabeta}; they have
the following numerical values

\begin{equation}
\begin{array}{l}
c_0 \equiv 39 956.8 \, \mathrm{cm} \, \mathrm{s}^{-1} \, ,
\vspace{0.25 cm}\\
\Omega_0  \equiv  8.09 \; \times 10^{-13}\; \mathrm{rad} \, \mathrm{s}^{-1} \, .
\end{array}
\label{velcsyomega}
\end{equation}
\noindent Then the initial
velocity of the $i$th SPH particle is
$ \vec{v}_i= \vec{\Omega_0} \times \vec{r}_i = \Omega_0(-y_i,x_i,0)\,$.

Besides, we implement a well known spectrum of
density perturbations on the initial particle distribution,
so that at the end of the simulation they might result in the formation
of binary systems. The evolution of a uniform density core like the model 
we consider
here has been reproduced by many groups worldwide using different
codes based both on grids and particles, see for instance
\cite{bodeniv}, \cite{sigalotti2001a} and \cite{sigalotti2001b} and the 
references there in. We
have also successfully reproduced the uniform model results in
\cite{NuestroApJ,NuestroRMAA}, from which we established the
reliability of our calculations for the evolution of the core using the
GAGDET2 code, as we followed the collapse process until peak
densities of the order of $2.9 \times 10^{-10} \, $ g $\, $cm$^{-3}$.

Let us say something about the gas thermodynamics. As the observed
star forming regions basically consist of molecular hydrogen cores
at $10\,$ K with an average density of $1 \times 10^{-20} \, $ g
$\,$ cm $^{-3}$, the ideal equation of state is a good approximation.
However, once gravity has produced a substantial core
contraction, the gas begins to heat. In order to take into account
this increase in temperature, we use the barotropic equation of
state proposed by \cite{boss2000}. Thus, in this paper we carry out
all our simulations using the barotropic equation of state

\begin{equation}
p= c_0^2 \, \rho \left[ 1 + \left(
\frac{\rho}{\rho_{crit}}\right)^{\gamma -1 } \, \right] ,
\label{beos}
\end{equation}
\noindent where for molecular hydrogen gas the ratio of specific
heats $\gamma$ is given by $\gamma\, \equiv 5/3$, because we only
take into account the translational
degrees of freedom of the hydrogen molecule. The most important free 
parameter is then
the critical density $\rho_{crit}$, which determines the change of thermodynamic
regime. Here we consider only one value of
$\rho_{crit}= 5.0 \times 10^{-14} \, $ g $\,$ cm$^{-3}$.


By comparing the results of our previous papers \cite{NuestroApJ,
NuestroRMAA} with those of \cite{whitehouse} for the collapse of an
isolated and rigidly rotating core with a uniform density profile,
we have concluded that the barotropic equation of state in general
behaves quite well and that it captures all the essential
thermodynamical phases of the collapse, see also
\cite{NuestroPlummer}. However, this does not mean that the same
comparison would be always correct for more complex initial
conditions or for collision models such as those we consider in
this paper.

Despite these facts and that we know that it is indispensable to
include all the detailed physics of the thermal transition in order
to achieve the correct results, be it in a collision or not, we
carry out the present simulations with the barotropic approximation,
because we know that there are other computational and physical
factors that could have a stronger influence on the outcome of a
simulation.

The collision cores considered for this work have pre-collision
velocities of up to several times the speed of sound, see Section
\ref{sec:modelos} and Table \ref{tab:modelos}. For these high
velocities a particle agglomeration is formed in the interface between 
the cores and
the artificial viscosity used for the calculations transforms kinetic
into thermal energy, which increases the temperature of the gas. However, as
we will show in Section~\ref{subsec:collisionprocess}, the cooling
time $\tau_{cool}$ is both much shorter than the local sound
crossing time, and the free fall time, hence the isothermal condition can
be used for the interaction region.

\section{Collision models and computational considerations}
\label{sec:modelos}

\subsection{The pre-collision geometry}
\label{subs:geo}

The simulations considered in this paper include both head-on and
oblique collisions. For the head-on collisions, the initial position
in cartesian coordinates of the center of mass (CM) of one
colliding cores is at $(x,y,z)=\vec{X}_{CM_1}=(0,R_0,R_0)$
and the other at $\vec{X}_{CM_2}=(0,-R_0,R_0)$. Recall that $R_0$
is the initial core's radius. The oblique
collisions are characterized by the magnitude and sign of an impact
parameter $b$, which depends on $R_0$. For the $b>0$ cases, the
colliding cores are initially displaced along the $X$-axis. For the
opposite case, when $b<0$, the cores are displaced again along the $X$-axis
but in the opposite direction. In all cases the displacement
is perpendicular to the symmetry axis of the collision (the $Y$-axis).

\begin{figure}
\begin{center}
\begin{tabular}{cc}
\includegraphics[width=1.5 in]{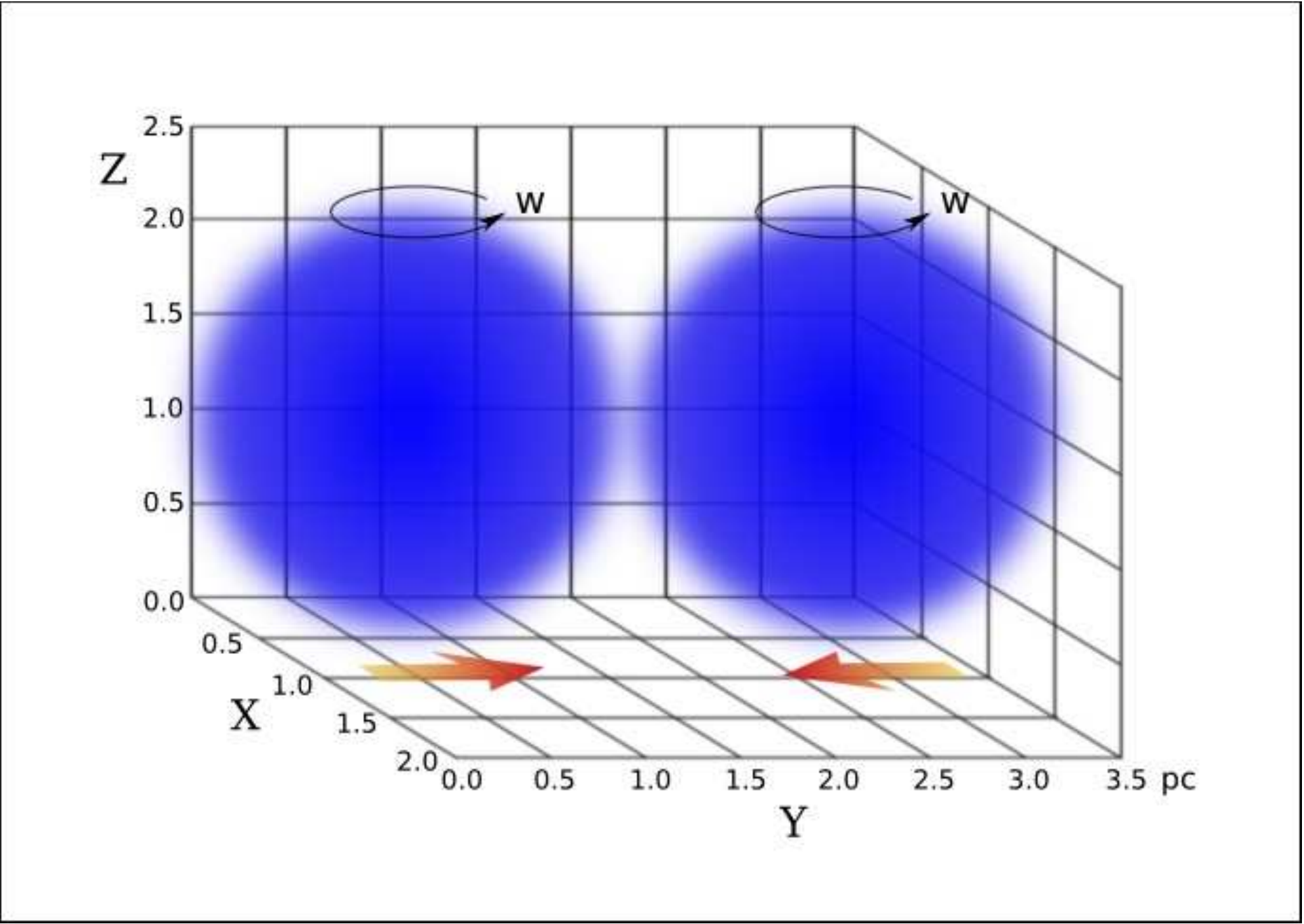} & 
\includegraphics[width=1.5 in]{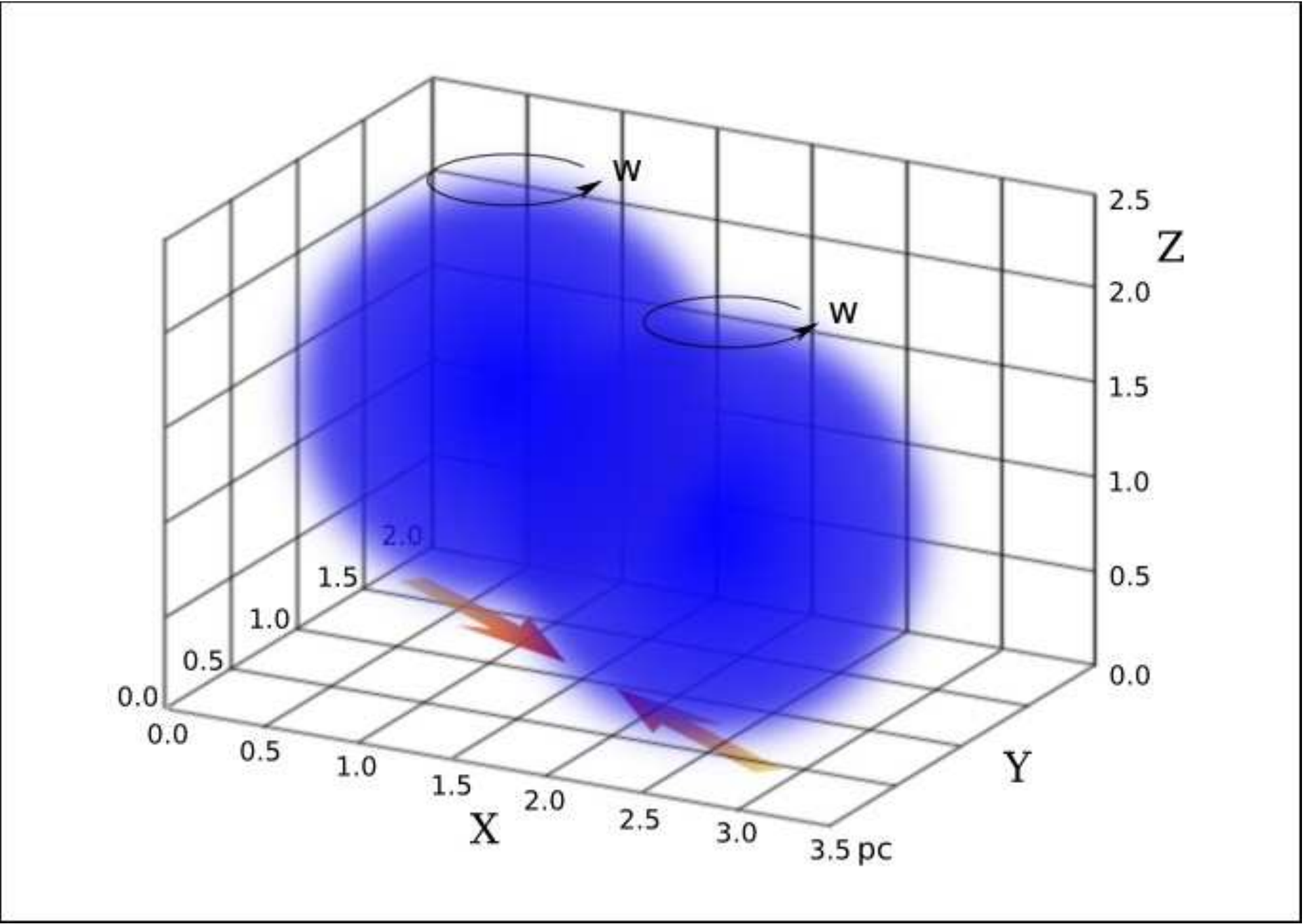}
\end{tabular}
\caption{\label{geometry} The initial geometry of the colliding cores for
head-on collisions (left panel), and for oblique collisions with a
positive impact parameter $b$ (right panel).} 
\end{center}
\end{figure}

For instance, for model $M1$ we place the first core so
that its center of mass in cartesian coordinates is
$\vec{X}_{CM_1}=(R_0/4,R_0,R_0)$, while the center of mass of the
second core is $\vec{X}_{CM_2}=(-R_0/4,-R_0,R_0)$. We consider this
orientation as a positive impact parameter $b$, see
Table~\ref{tab:modelos} and Fig. \ref{geometry}. Analogously, for
model $M3$, the center of mass of the cores is placed at the
coordinates $\vec{X}_{CM_1}=(-b/2,R_0,R_0)$ and
$\vec{X}_{CM_2}=(b/2,-R_0,R_0)$, which corresponds to a negative
impact parameter.

For the collision models we consider four different impact
parameter values $b=0,R_0/4,\pm \, R_0/2$, and pre-collision
velocities in the range of 0.23 to 6.4 times the speed of sound. It should
be noticed that this range of values is well motivated physically,
as it follows from a statistical study of core collision in the
context of interacting galaxies, see \cite{bekki}.


\subsection{The pre-collision velocity}
\label{subs:mod}

In this paper, we define the pre-collision velocity in the same way
for both head-on and oblique collisions, as follows. The average
velocity of the center of mass of the cores points along the
$Y$-axis, and are $\vec{V}_{CM_1}=(0,-V_{ave},0)$ and
$\vec{V}_{CM_2}=(0,V_{ave},0)$. We define the pre-collision speed
$V_{app}$ (with respect to the center of mass of the colliding
binary system) of the cores by means of $V_{app}= 2 \, V_{ave}$.
Therefore, the pre-collision velocity  $V_{app}$ of the colliding
cores, also known as the relative or impact velocity of the cores,
should be in the range $0.30-20$ Mach.

However, there must be a correlation between the pre-collision
velocity of the colliding cores and their separation. As we assume
here that the cores are initially in close contact, then we can not
go to arbitrarily large values of the pre-collision velocities.
Hence the values used here for $V_{app}$ have been chosen to
represent an ample range, where the collision timescale can be
compared with the free-fall time scale, as discussed in Section
\ref{sec:dis}.
\subsection{The Evolution Code}
\label{subs:code}

We carry out the time evolution of the initial distribution of
particles with the fully parallel GAGDET2 code, which is described
in detail by \cite{gadget2}. GADGET2 is based on the tree-PM method
for computing the gravitational forces and on the standard SPH
method for solving the Euler equations of hydrodynamics. GADGET2
incorporates the following standard features: (i) each particle $i$
has its own smoothing length $h_i$; (ii) the particles are also
allowed to have an individual gravitational softening length
$\epsilon_i$, whose value is adjusted such that for every time step,
$\epsilon_i \, h_i$ is of order unity. GADGET2 fixes the value of
$\epsilon_i$ for each time-step using the minimum value of the
smoothing length of all particles, that is, if $h_{min}=min(h_i)$
for $i=1,2...N$, then $\epsilon_i=h_{min}$.

The GADGET2 code has an implementation of a Monaghan-Balsara form
for the artificial viscosity, see~\cite{mona1983} and
\cite{balsara1995}. The strength of the viscosity is regulated by
setting the parameter $\alpha_{\nu} = 0.75$ and $\beta_{\nu}=\frac{3}{2}\,
\times \alpha_{\nu}$, see Equation (14) in~\cite{gadget2}. We fix
the Courant factor to $0.1$. The number of smoothing neighbour particles
is fixed to 40 and the allowed variation is fixed to 5 particles, so that 
the GADGET2 will
adjust the kernel's smoothing length such that the number of neighbours is
always kept within the range 40 $\pm$ 5.

\subsection{The detection of accretion centers}
\label{subsec:acccent}

Let us now describe the modification implemented into the GADGET2
code to detect accretion centers. Any gas particle with density
higher than $\rho_s$ is a candidate to be an accretion center. We
localize all the candidate particles for a given time $t$. We then
test the separation between candidate particles: if there is one
candidate with no other candidate closer than $10\times r_{int}$,
then this particle is identified as an accretion center at time $t$.
We define $r_{int}$ as the neighbor radius for an accretion center,
given by $r_{int}=1.5\times h_{min}$. In this way $r_{int}$
determines a set of particles which are within the sphere having
this radius and whose center is the accretion center itself. All
those particles will give their mass and momentum to the accretion
center. We change the GADGET2 {\it particle type} for all those
particles within the neighbor radius from being $0$ to $-1$, which is
a particle type undefined in GADGET2 and therefore they will be not advanced
in time anymore.

\begin{figure}
\begin{center}
\begin{tabular}{cc}
\includegraphics[width=2.5in]{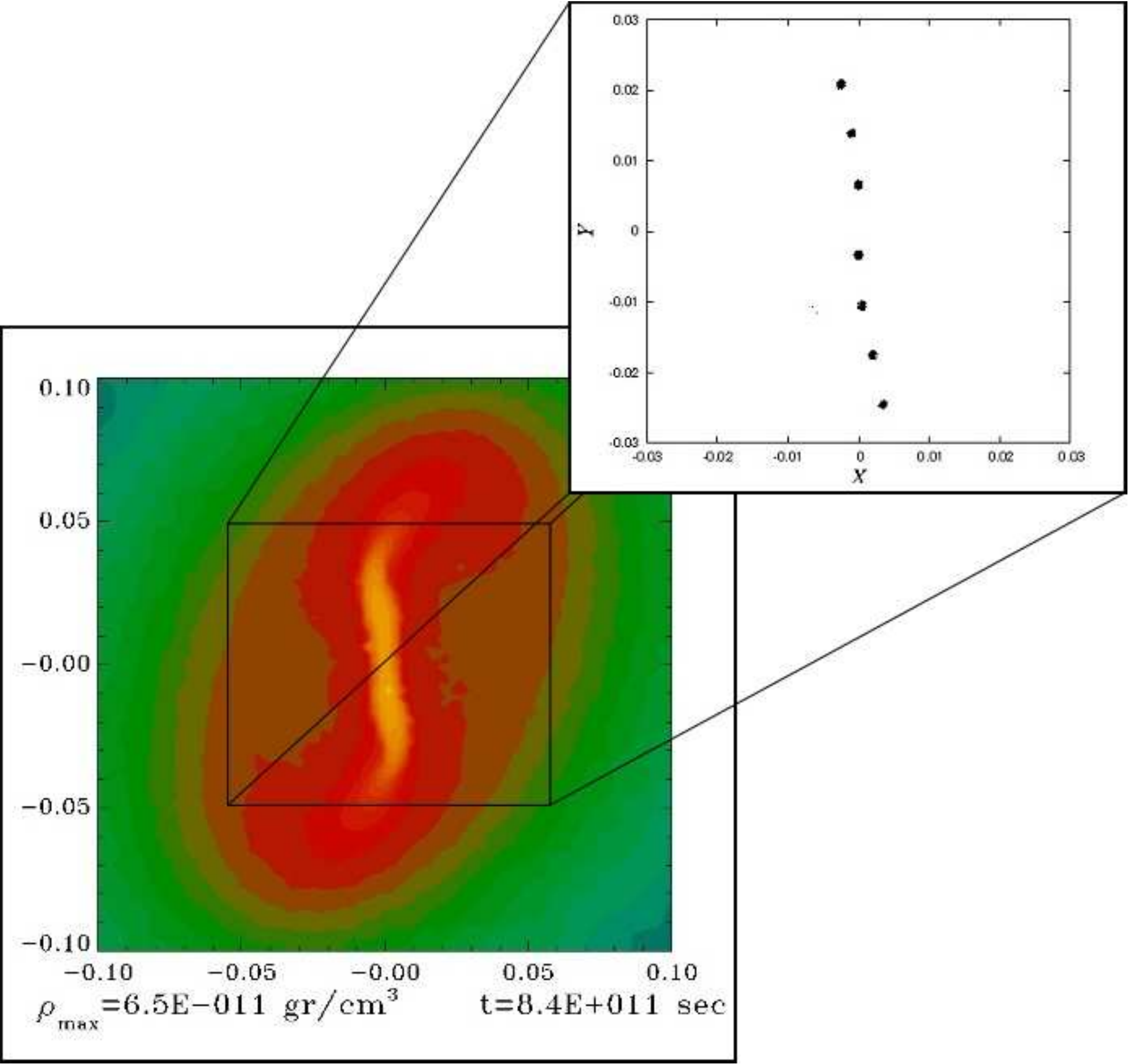} &
\includegraphics[width=2.5in]{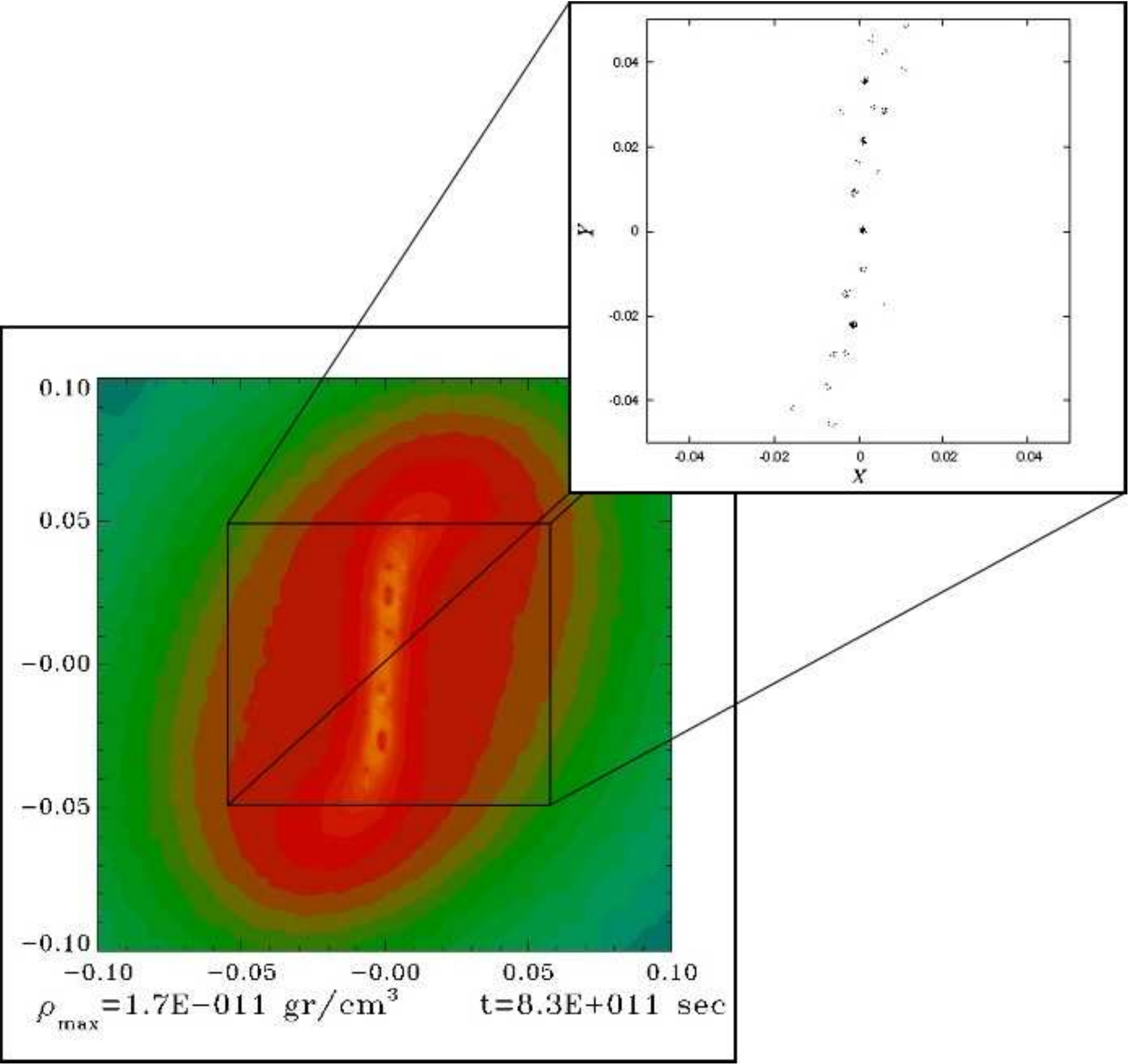}
\end{tabular}
\caption{ \label{modMCmR} The last snapshots available for the collapse of the
isolated core, for which we use the modified GADGET2 with density
cutoff given by $\rho_s = 1.0 \times 10^{-12} \, $ g $\, $ cm$^{-3}$
(left panel), and $\rho_s = 5.0 \times 10^{-14}\; $g $\, $cm$^{-3}$
(right panel).}
\end{center}
\end{figure}

We go on to determine an appropriate value of $\rho_s$ by using a
uniform and rotating core model studied in previous collapse
calculations. When we use a low value for $\rho_s$, for instance,
$\rho_s= 5.0\times 10^{-14} \, $g cm$^{-3}$, then too many
accretion centers are formed even in an early collapse stage, see
the right panel of Fig.~\ref{modMCmR}. A better value for $\rho_s$
would be $\rho_s= 1.0\times 10^{-12} \, $ g cm$^{-3}$, as the
evolution of the test model both with and without the implemented
code are quite similar. 

The dependence of the number of accretion centers  
on the parameter $\rho_s$ can be considered as a depending way of 
{\it discretize} the densest gas region of interest. We emphasize that 
the radius $r_{int}$ and the $\rho_s$ are chosen here such that 
the formation of accretion centers does not 
affect the subsequent dynamics of the gas outside their influence region, 
otherwise    
it would be obviously a big problem in our simulations. 

Furthermore, we mention here that \cite{klessen} have used  a clump finding 
algorithm, fully integrated into the SPH formalism, which makes no use of any 
parameters like 
ours $\rho_s$ and $r_{int}$. 

However, other methods to detect sink particles like the 
one described by \cite{bate} depends on three parameters: (i) a density 
threshold (which was chosen $10^5$ times higher than the average cloud density) 
to initially point to the gas candidate to be a sink particle; (ii) an 
accretion radius and 
(iii) an inner radius (which was defined to be 10-100 times smaller than 
	the accretion radius).    

\cite{bate} applied the sink particle technique to study the
collapse of the so called standard isotermal test case.  Their objective was  
to extend the collapse calculation beyond the protostar formation by removing 
all those particles in the densest regions, which have very small time steps. 

\cite{bate} applied several collapsing and bounding tests on the gas particles 
inside 
the accretion radius before they are decided to be accreted or not by the sink 
particle. Besides, all those particles inside the inner radius are 
accreted by the sink particle regarless of the tests. 

We must clarify that our parameter $r_{int}$ plays  
here the role of the {\it inner accretion radius} defined by \cite{bate} and 
this 
is why we decide to skip all the tests as well. 
In a forthcoming revision of our code we will include an exterior radius 
$r_{ext}$ 
in order to detect and eventually eliminate all those particles located 
within an exterior  
influence region of an accretion center.

\cite{fede} implemented this sink idea in the mesh based code FLASH, which uses 
an adaptive mesh refiment (AMR) technique. They defined a sink cell and a sink 
volume by introducing two basic parameters: a density threshold and a radius. 
They went on to apply 6 tests on the sink cell to avoid the formation of 
spurious sinks. They found good agrement between this implementation 
and that reported by \cite{bate} in the case of the core collapse. However, 
they emphasized that the use a sole density threshold parameter for deciding 
sink creation is insufficient when supersonic shocks are present. 

\begin{figure}
\begin{center}
\begin{tabular}{c}
\includegraphics[width=3.0in]{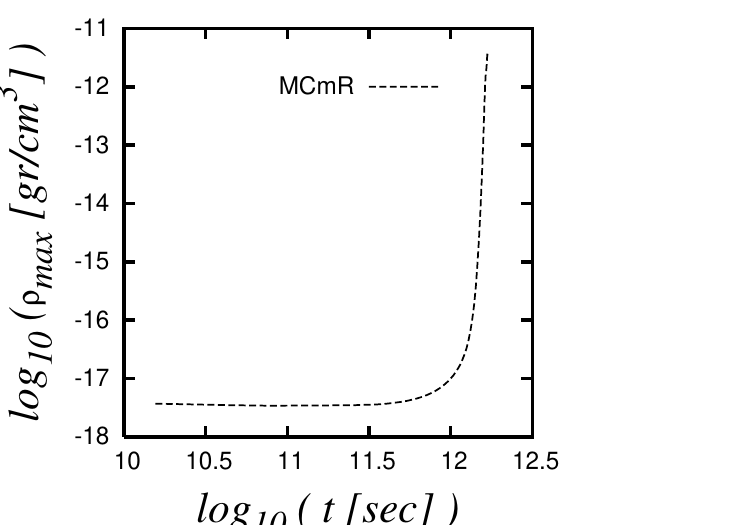} \\
\includegraphics[width=3.0in]{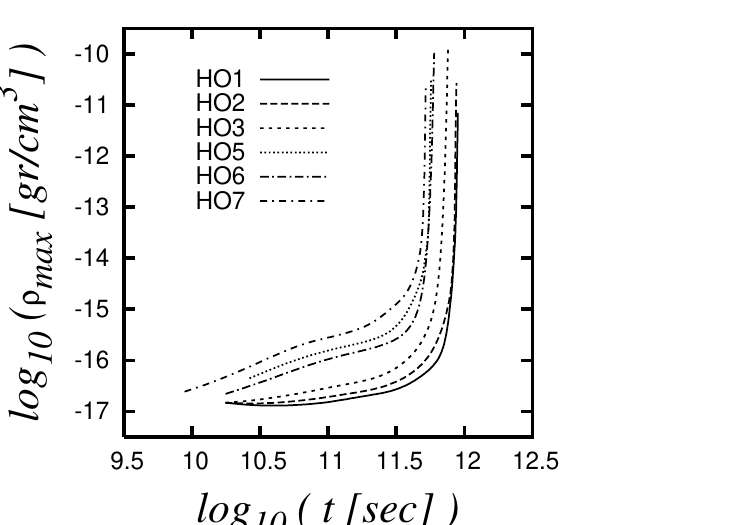} \\
\includegraphics[width=3.0in]{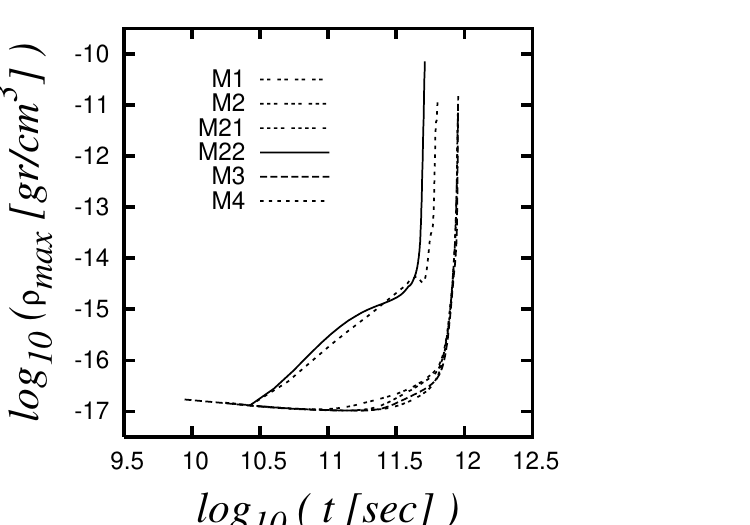}
\end{tabular}
\caption{ \label{denmax} Time evolution of the peak density for: the isolated
core (top panel), the head-on collision models (middle panel), and
 the oblique collision models (bottom panel).} 
\end{center}
\end{figure}

Despite of all this, we carry out the time evolution of all the
initial distribution of particles for the colliding models with a
fixed value of $\rho_s= 5.0\times 10^{-14} \,$ g cm$^{-3}$, because of
the failure to evolve many of the particles of the colliding models until
densities higher than $\rho_s$, otherwise the time step of the GADGET2
code becomes very small when not many particles are able to reach densities
beyond $\rho_s$. As expected, all the collision models show a strong
tendency to collapse like the isolated core, as can be seen in
Fig.~\ref{denmax}.  The application of the limited technique described in 
this Section makes sense for us, as the main aim of this paper is only to 
detect the places where  the densest region are formed when two 
rotating cores collide with a moderate pre-collision velocity.

\subsection{Resolution}
\label{subs:resol}

According to \cite{truelove}, the resolution requirement for
avoiding the growth of numerical instabilities is expressed in terms
of the Jeans wavelength $\lambda_J$, which is given by

\begin{equation}
\lambda_J=\sqrt{ \frac{\pi \, c^2}{G\, \rho}} \; , 
\label{ljeans}
\end{equation}
\noindent where $G$ is Newton's universal gravitation constant, $c$
is the instantaneous sound speed and $\rho$ is the local density. To
obtain a more useful form for a particle based code, the Jeans
wavelength $\lambda_J$ is written in terms of the spherical Jeans
mass $M_J$, which is defined by

\begin{equation}
M_J \equiv \frac{4}{3}\pi \; \rho \left(\frac{ \lambda_J}{2}
\right)^3 = \frac{ \pi^\frac{5}{2} }{6} \frac{c^3}{ \sqrt{G^3 \,
\rho} } \;. \label{mjeans}
\end{equation}

Nowadays it is well known that the Jeans requirement $l<\lambda_J/4$
(where $l$ is a characteristic length scale of the grid for a mesh
based code) is a necessary condition to avoid the occurrence of
artificial fragmentation. For particle based codes,
\cite{bateburkert97} showed that there is also a {\it mass limit
resolution criterion} which needs to be fulfilled besides that of
\cite{truelove}. They showed that an SPH code produces correct
results involving self-gravity as long as the minimum resolvable
mass is always less than the Jeans mass $M_J$.

Now, if we approximate the instantaneous sound speed by $c=\sqrt{p /
\rho}$, then according to Eq. \ref{beos}, we have that
\begin{equation}
M_J = \frac{ \pi^{ \frac{5}{2} } }{6} \; \frac{c_0^3}{\sqrt{G^3\,
\rho} } \; \left[1+ \gamma \, \left( \frac{\rho}{\rho_{crit}
}\right)^{\gamma-1}\right]^{\frac{3}{2}} \; .
\label{mjeanse}
\end{equation}

Following \cite{bateburkert97}, the smallest mass that an SPH
calculation can resolve is $m_r \approx M_J / (2 N_{neigh})$, where
$N_{neigh}$ is the number of neighbor particles included in the
SPH kernel. For our collision models to comply with the Jeans
requirement, the particle mass $m_p$ must be such that $m_p/m_r<1$.

In this work, we set $N=1,000,000 \;$ SPH particles
for representing the initial core configuration in each model. By
means of a rectangular mesh we make the partition of the simulation
volume in small elements each with a volume $\Delta x\, \Delta y\,
\Delta z $; at the center of each volume we place a particle -the
$i-th$, say- with a mass determined by its location according to the
density profile being considered, that is: $m_i= \rho(x_i,y_i,z_i)
\, \Delta x\, \Delta y\, \Delta z$ with $i=1,...,N$. Next, we
displace each particle from its location a distance of the order
$\Delta x/4.0$ in a random spatial direction. 

For verifying that the Jeans stability condition is satisfied we
will use the collision model $HO-3$, which is the one that reached
the highest maximum density of all models, see Table
\ref{tab:modelos}. For this model we follow its collapse until a
peak density of $\rho_{max}=2.8 \times 10^{-10} \; $g cm$^{-3}$.
The particle mass is $m_p \equiv 2 \, M_0 / N_p = 8 \times 10^{-6}
M_{\odot}$, where $M_t=16 M_{\odot}$ is the total mass contained in
the simulation box and $N_p$ is the total number of particles, which
are $2$ million for all the models in this paper.

The minimum Jeans mass for model $HO-3$ is given by
$\left(M_J\right)_{HO-3} \approx 8.6 \times 10^{-4} \; M_\odot$,
from which we obtain $m_r=1.0 \times 10^{-5} \, M_\odot$. Thus, for
model $HO-3$ we obtain the ratio $m_p/m_r=0.74$, and the Jeans
resolution requirement is satisfied very easily.

For the rest of the collision models, the minimum Jeans mass is
expected to be greater than for model $HO-3$ because their maximum
density is less than for model $HO-3$. It is then clear that for all
models the Jeans length criterion is satisfied as well.
\subsection{Collision models}
\label{subs:colimodels}


We summarize the collision models considered in this paper in Table
\ref{tab:modelos}, whose entries are as follows. Column one shows
the label chosen to identify the model. Column two indicates the
impact parameter $b$ of the collision models in terms of $R_0$;
note the appearance of a sign before the magnitude of
$b$, which is related to the orientation of the colliding cores
along the $X$-axis, see Section \ref{subs:geo}. Column three indicates
the pre-collision velocity of the colliding cores, expressed in terms
of the initial speed of sound in the core, whose numerical value is
given in Eq.~\ref{velcsyomega}. In the fourth column we show the
number $N_{acc}$ of accretion centers found for
each system, both for the case of an isolated core and for the colliding 
models. In
the fifth column we show the average number of SPH particles
$Np_{acc}$ per accretion center formed in each model while in the
sixth column we show average mass of the accretion center
$M_{av}/M_{sun}$. As a way of comparing our simulations with other
simulations, in the seventh column we show the peak density
$\rho_{max}$ reached in each run.
\section{Results}
\label{sec:resultados}

In order to show the results of our simulations, we present
iso-density plots for a slice of matter parallel to the $X-Y$ plane.
The procedure is as follows. We first locate the SPH particle with
the maximum density in the entire volume space of a simulation. The
$z$-coordinate of that particle, say $z_{max}$, determines the height
of the thin slice of material. The width of the slice is determined
so that about $10,000\, $SPH particles enter into the slice, which
is centered around the $z_{max}$ coordinate.

Once the SPH particles defining the slice have been selected, we
set a color scale related to the iso-density curves: yellow indicates
areas with higher densities, blue indicates those with lower
densities, and green and orange indicate intermediate density
regions. It should be noted that there is no relation between the
density colors associated with different panels even in the same
plot.  At the bottom of each iso-density panel, we include two numerical
values to illustrate the different stages of the evolution process:
the left value is the peak density $\rho(t)$ at time $t$; and the
right one is the actual time $t$ in seconds.


\subsection{The general picture of a core collision}
\label{subsec:collisionprocess}

\begin{figure}
\begin{center}
\begin{tabular}{cc}
\includegraphics[width=2.2in]{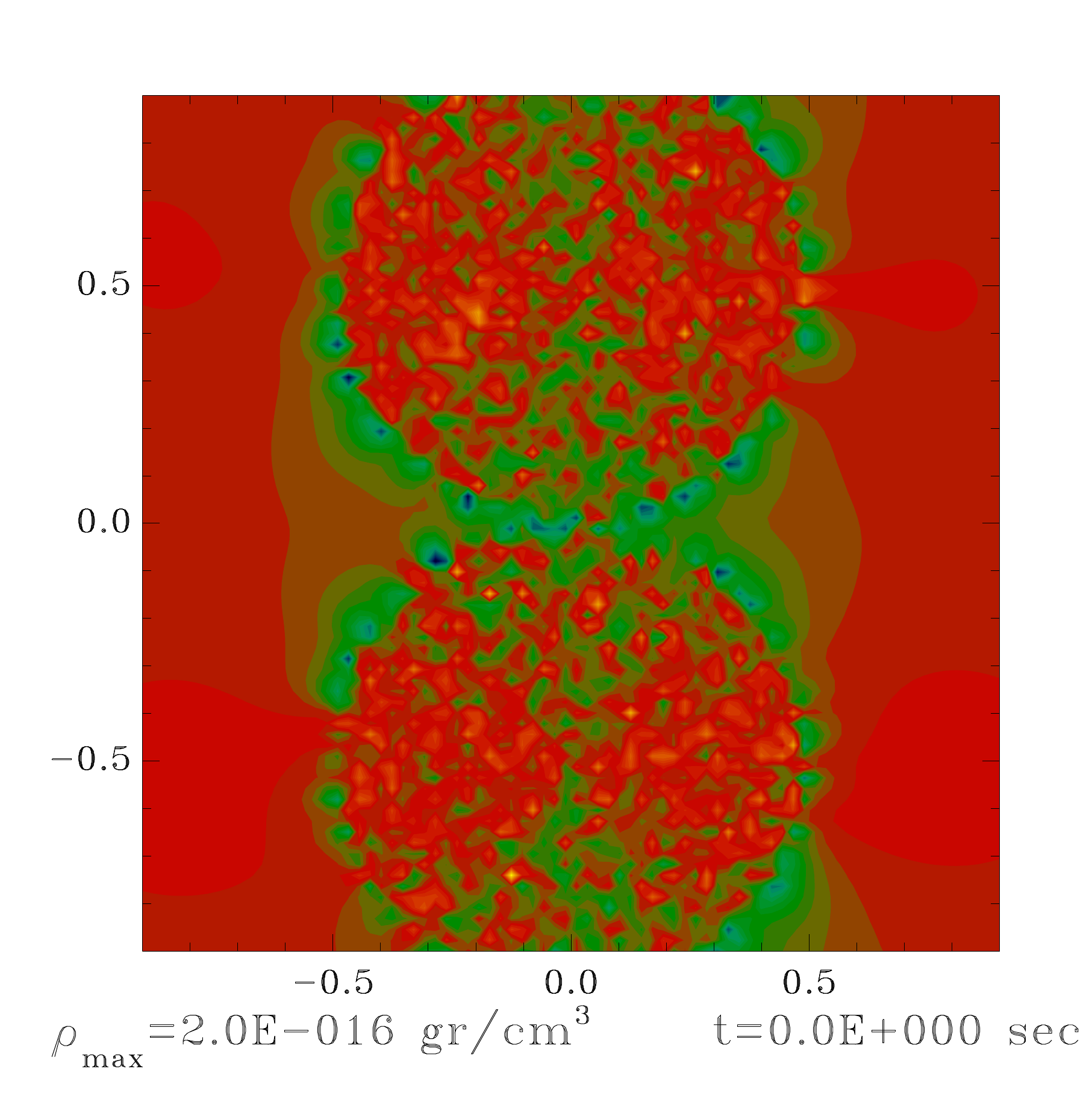} & 
\includegraphics[width=2.2in]{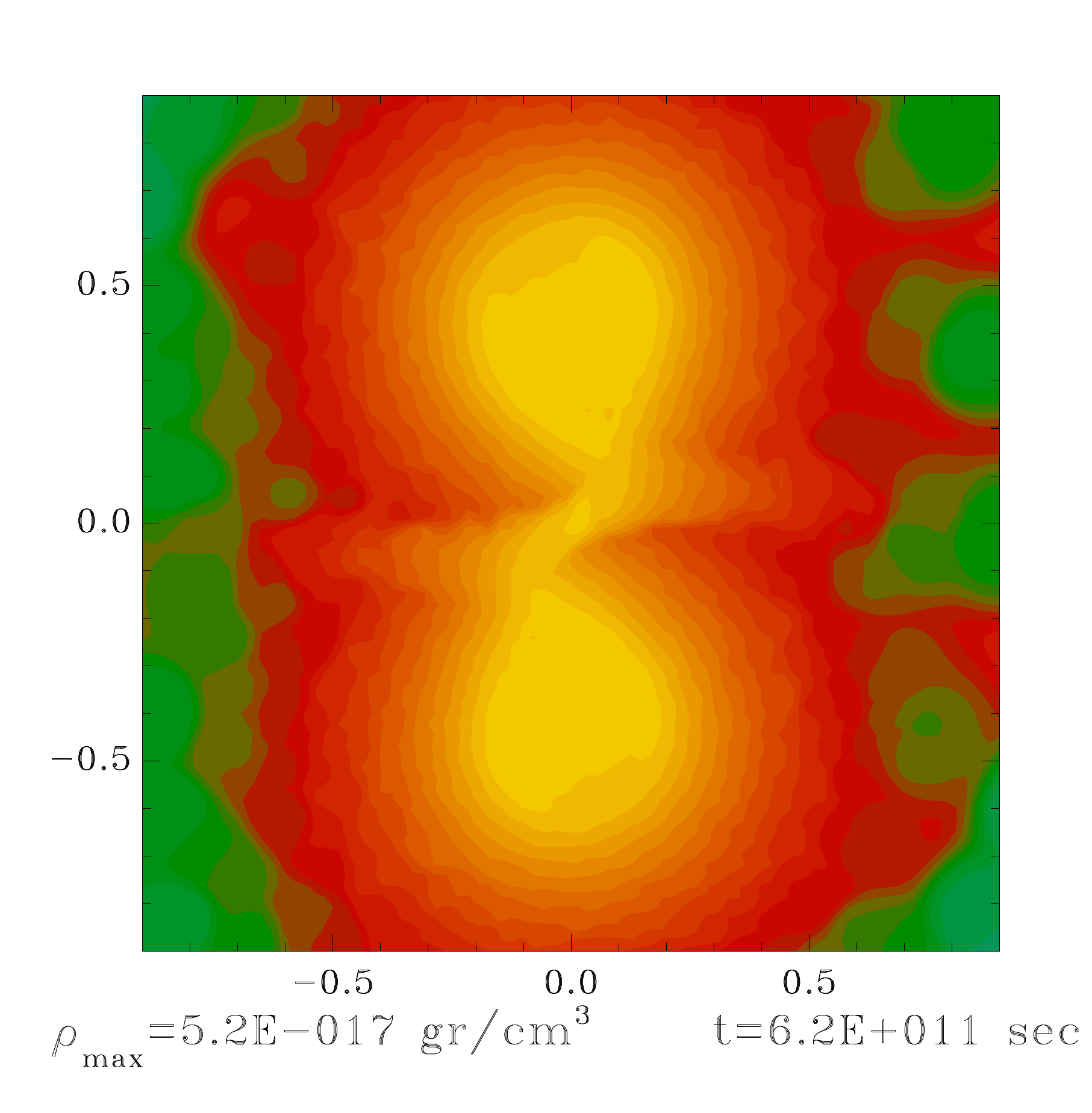} \\
\includegraphics[width=2.2in]{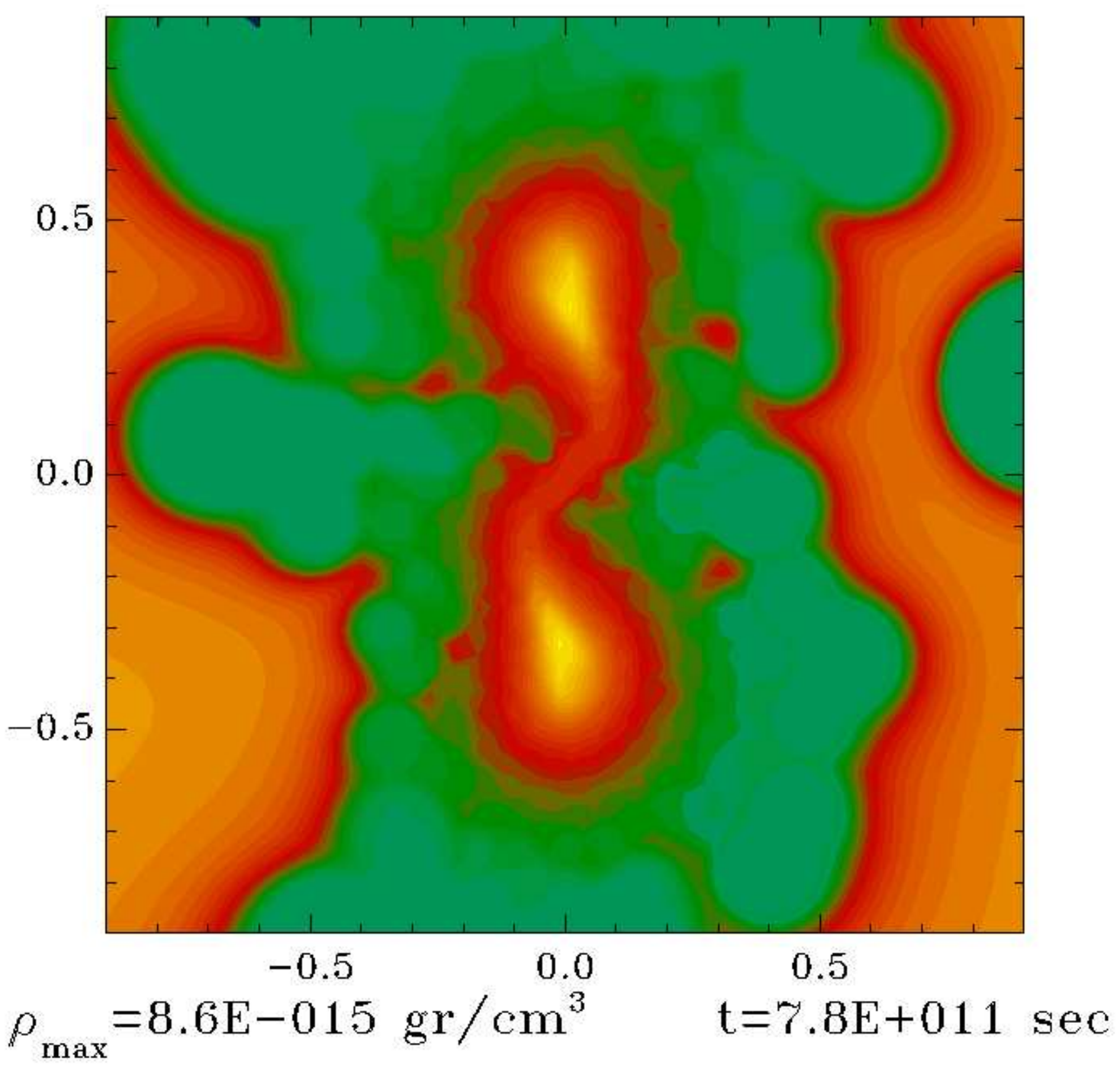} & 
\includegraphics[width=2.2in]{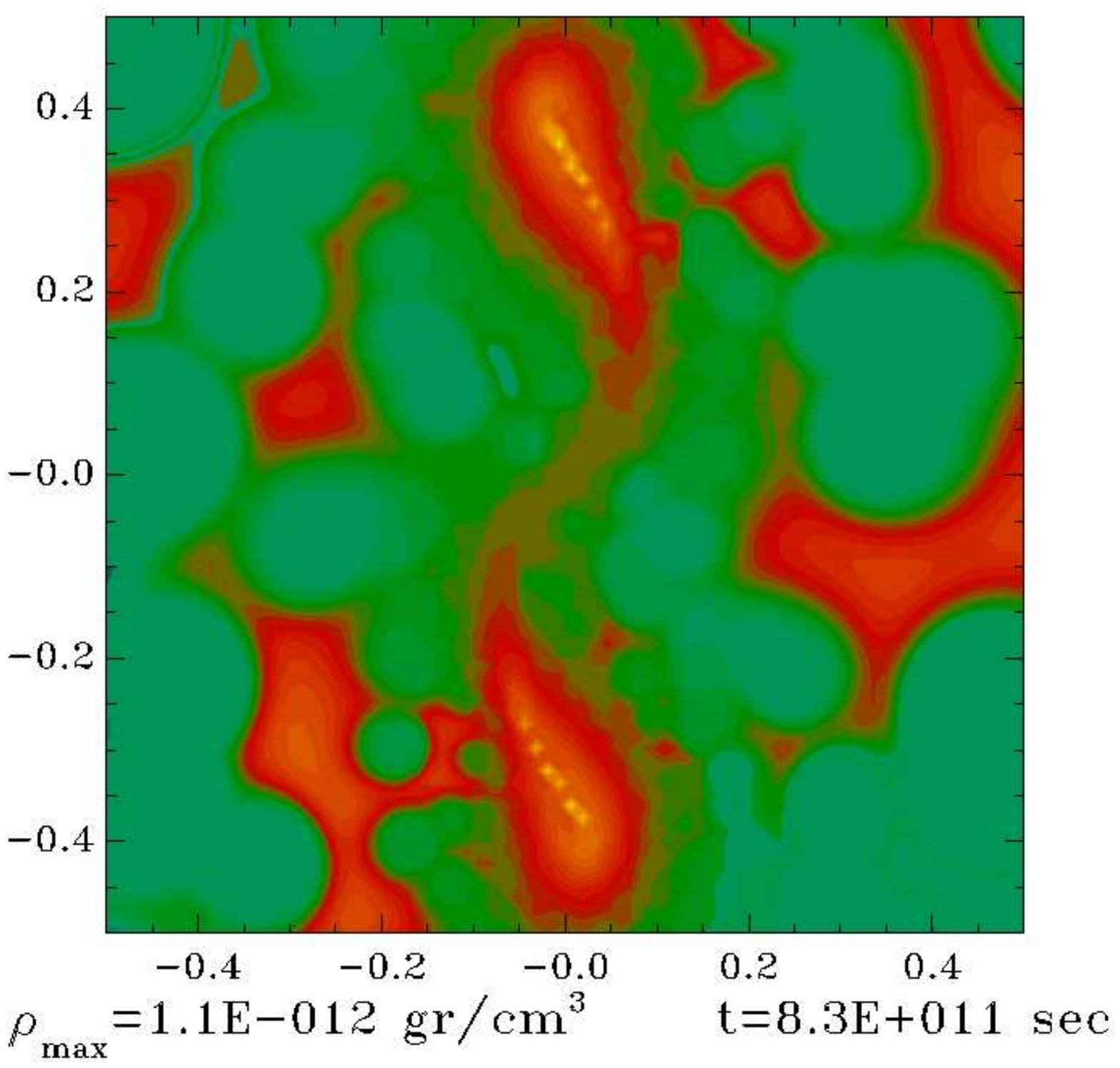}
\end{tabular}
\caption{ \label{HO1norgad} Iso-density plot illustrating the gravitational 
collapse process for model $HO1$ computed with the normal GADGET2 code.}
\end{center}
\end{figure}

Let us now briefly describe the general picture of a collision
system by using the evolution of model $HO-1$ as an example, calculated using
the normal GADGET2 code just for comparison. In the first top left panel of
Fig.~\ref{HO1norgad}, one can see that two identical cores are just placed one
against the other at $t=0$. Each core shares those particles which are
closer to the edge of the neighboring core.

As each core in the colliding system is collapsing separately as
well, we see in the right bottom panel of Fig.
\ref{HO1norgad} the formation of a filament in each
central core region. Such filaments are indeed connected by a gas bridge that
is formed in the interface region of the cores.

The interface particles are compressed and a front perturbation is
formed that propagates into the two cores. The artificial viscosity
then transforms kinetic energy into heat. We expect that the
barotropic equation of state proposed by \cite{boss2000} and
described in Eq.~\ref{beos} mimics the radiative cooling needed as
this heat must be radiated away in a very short timescale, so that
we can assume that the cores and the interface region between the
cores remain isothermal. The perturbation wave that propagates into
the cores can disturb the filaments in the central region.

During the collision process a slab of material is formed along the
collision front. The cores then expand but eventually collapse
again. The compressed slab formed is susceptible of having various
instabilities, which have been studied by \cite{vishniac1983} for
the linear regime, and by \cite{vishniac1994} for the non-linear
case. For the present work, the relevant instabilities are the
shearing and the gravitational. The shearing instability dominates
at low density while the gravitational one takes over for densities
above the critical density

\begin{equation}
\rho_{shear} \, \sim 7.86132 \, \times 10^{-22} \left (
\frac{\eta}{pc} \right ) \left ( \frac{\lambda_{NTSI}}{pc} \right )
\left ( \frac{T}{K} \right ) \, \mathrm{g} \, \mathrm{cm}^{-3}, 
\label{rhocrit}
\end{equation}
\noindent where $\eta$ is the amplitude of the perturbation,
$\lambda_{NTSI}$ the length of the unstable mode, and $T$ the
temperature, see \cite{heitsch08}.

Of the shearing instabilities, the main ones are the non-linear thin
shell instability (NTSI), and the Kelvin-Helmholtz (KH) instability.
The NTSI instability is expected to occur in the shocked slab just
after the core collision, and the non-linear bending and breathing
modes could also be present. From our numerical results we estimate
that the parameters in Eq. \ref{rhocrit} are $\lambda_{NTSI} \, \sim
\, 0.1 \,$ pc, $\eta \, \sim 2 \, \lambda_{NTSI}$ and $T \, \sim \,
10 \,$ K, and so the $NTSI$ and $KH$ instability are suppressed when
the density goes above $\rho_{crit} \, \sim 1.57 \, \times 10^{-18}
\, $g $\,$ cm$^{-3}$.


In the next sections we shall describe the results obtained with the
modified GADGET2 code for the time evolution of all models. The aim
is to locate accretion centers and to determine their properties along
the filaments and at the interfacial region as well.

\subsection{Head-on collisions}
\label{subsec:sinbrazo}

\begin{figure}
\begin{center}
\begin{tabular}{ccc}
\includegraphics[width=2.2in]{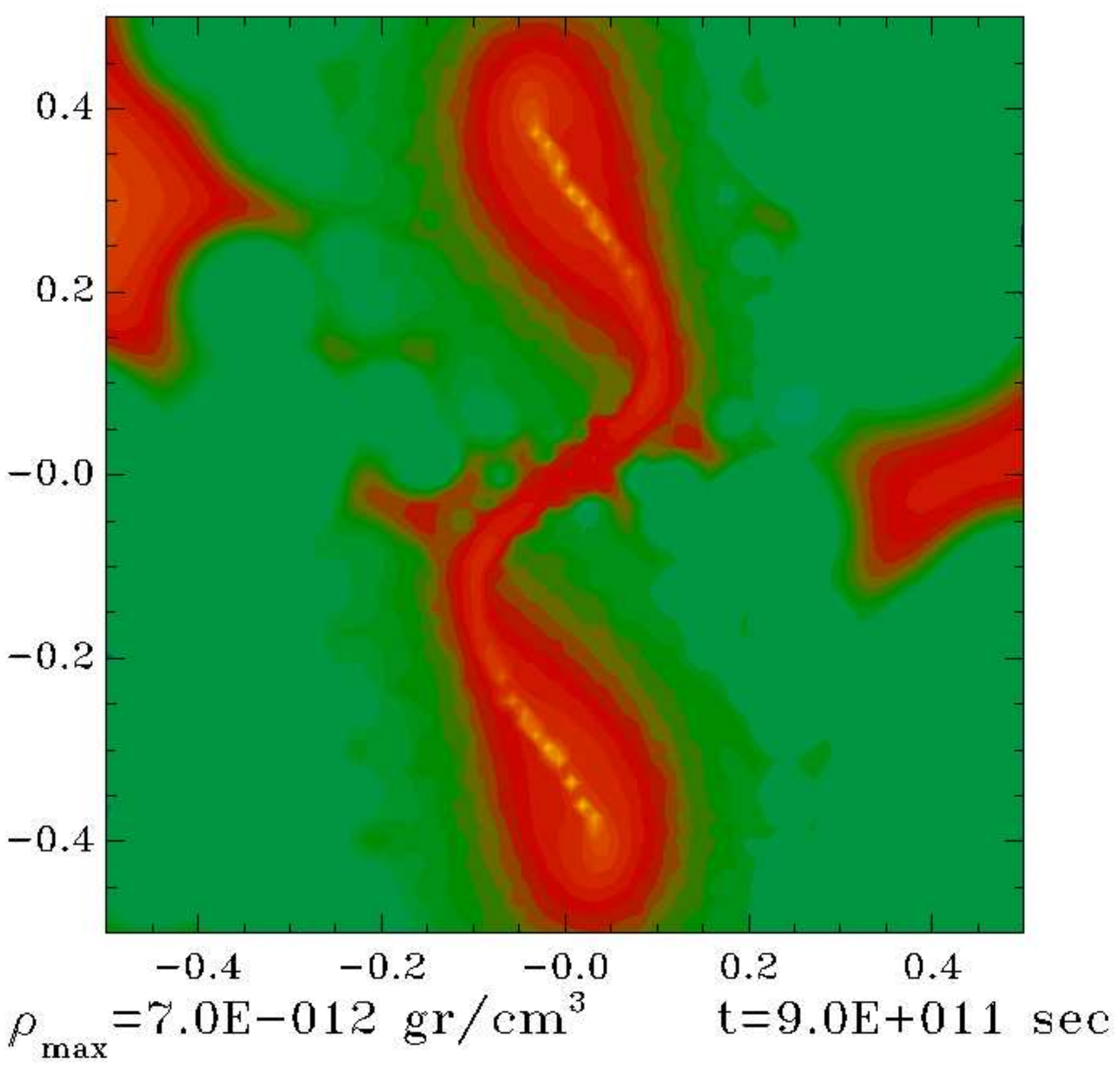} & 
\includegraphics[width=2.2in]{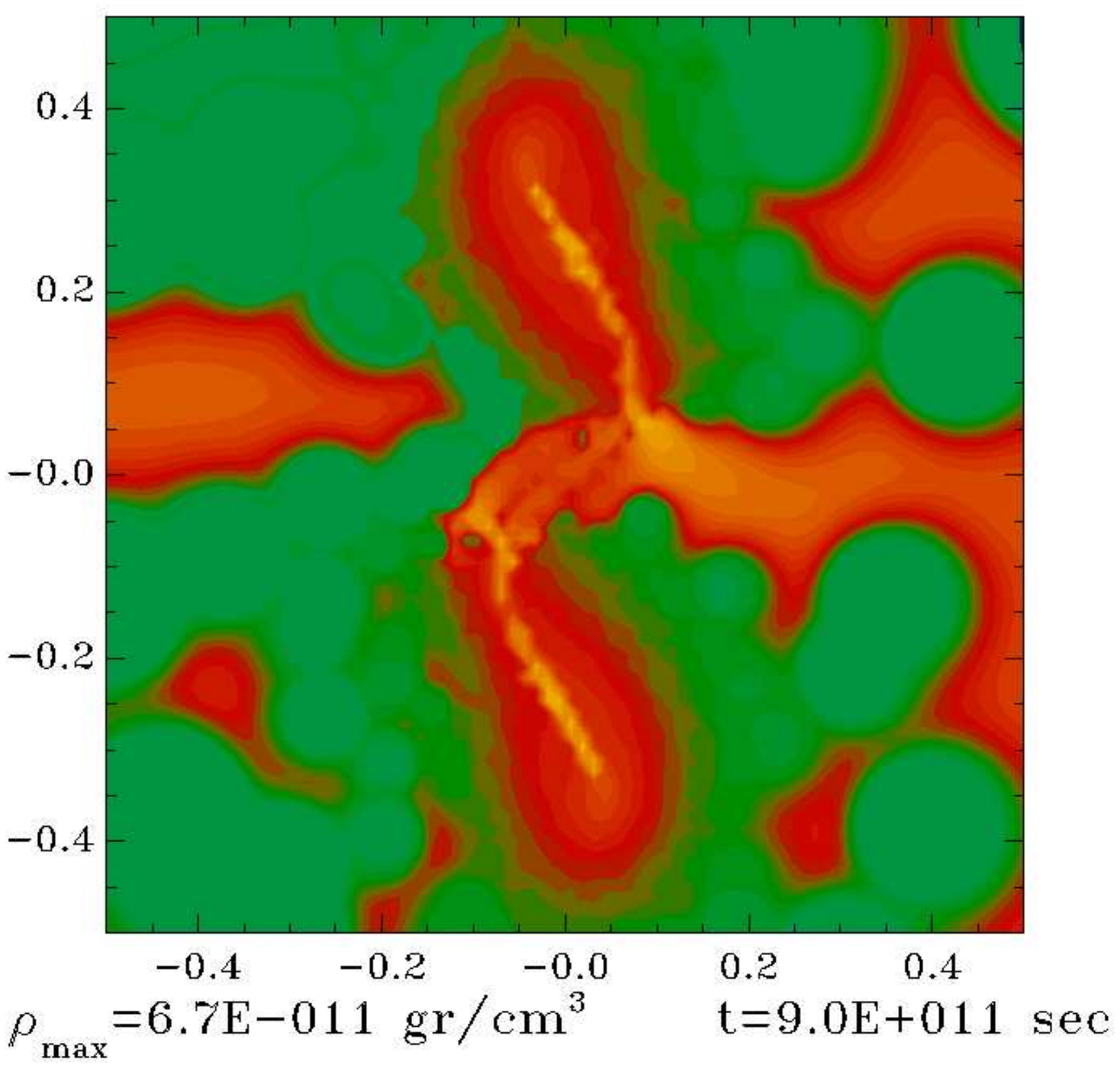}
& \includegraphics[width=2.2in]{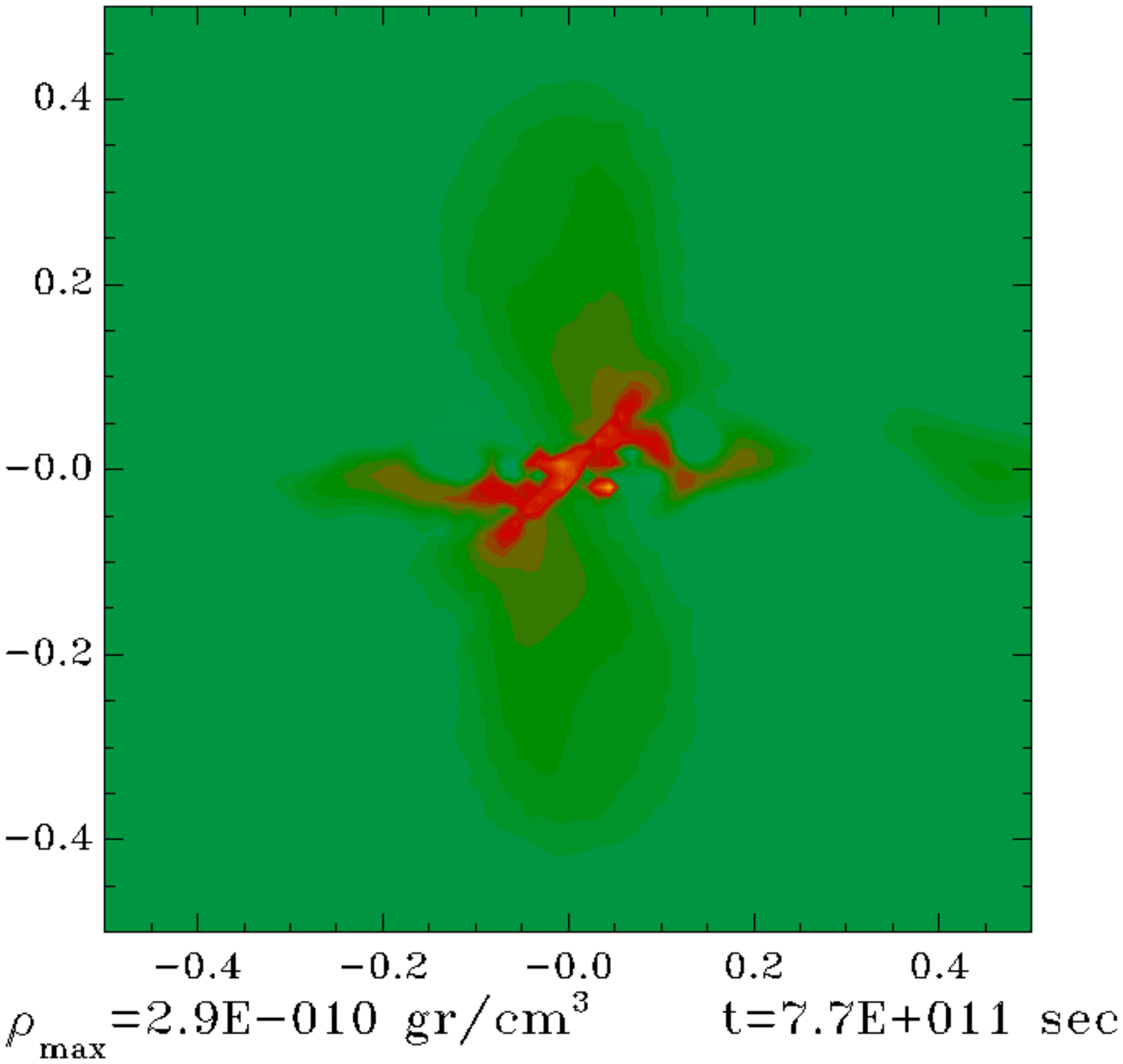} \\
\includegraphics[width=2.2in]{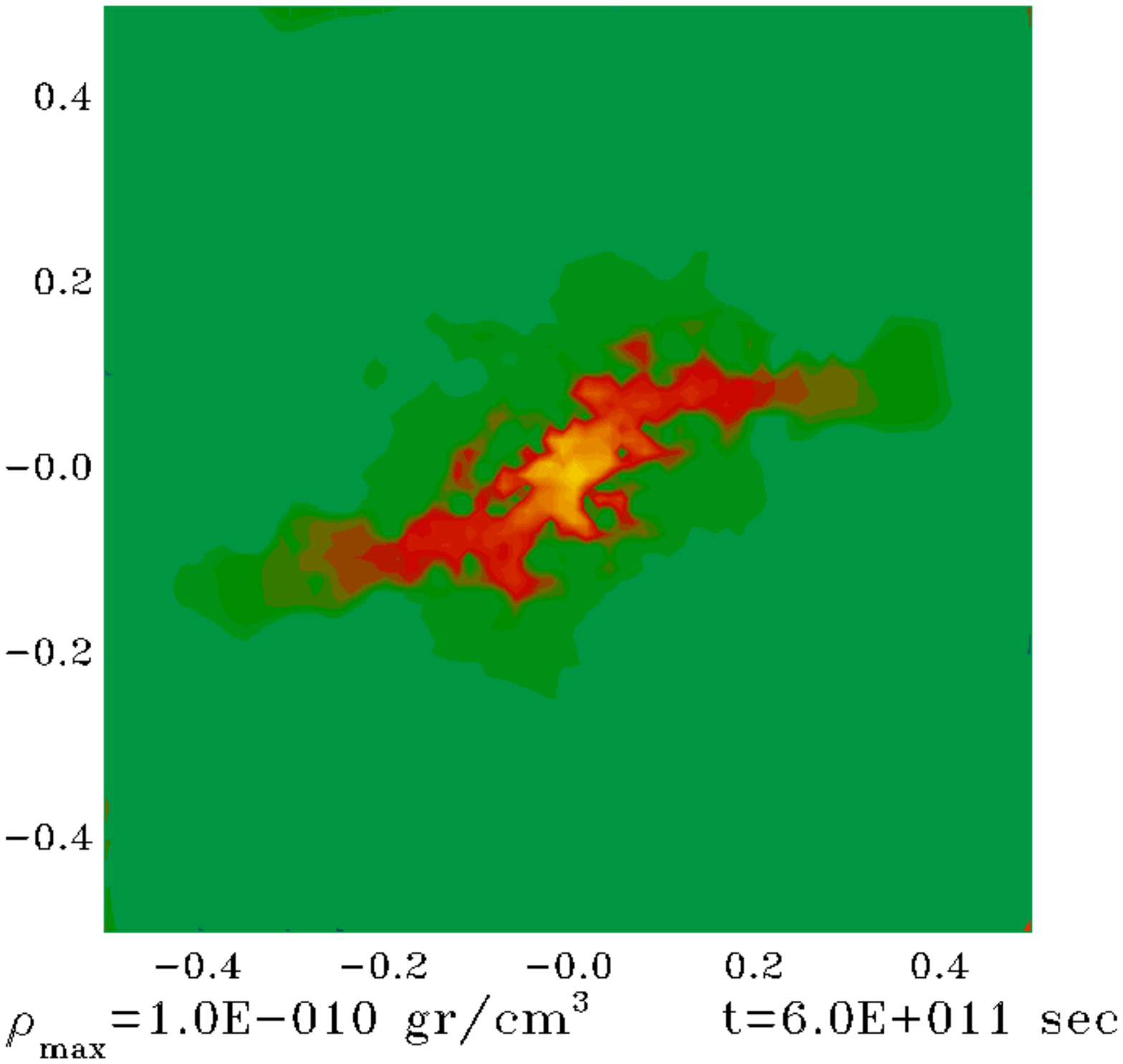} & 
\includegraphics[width=2.2in]{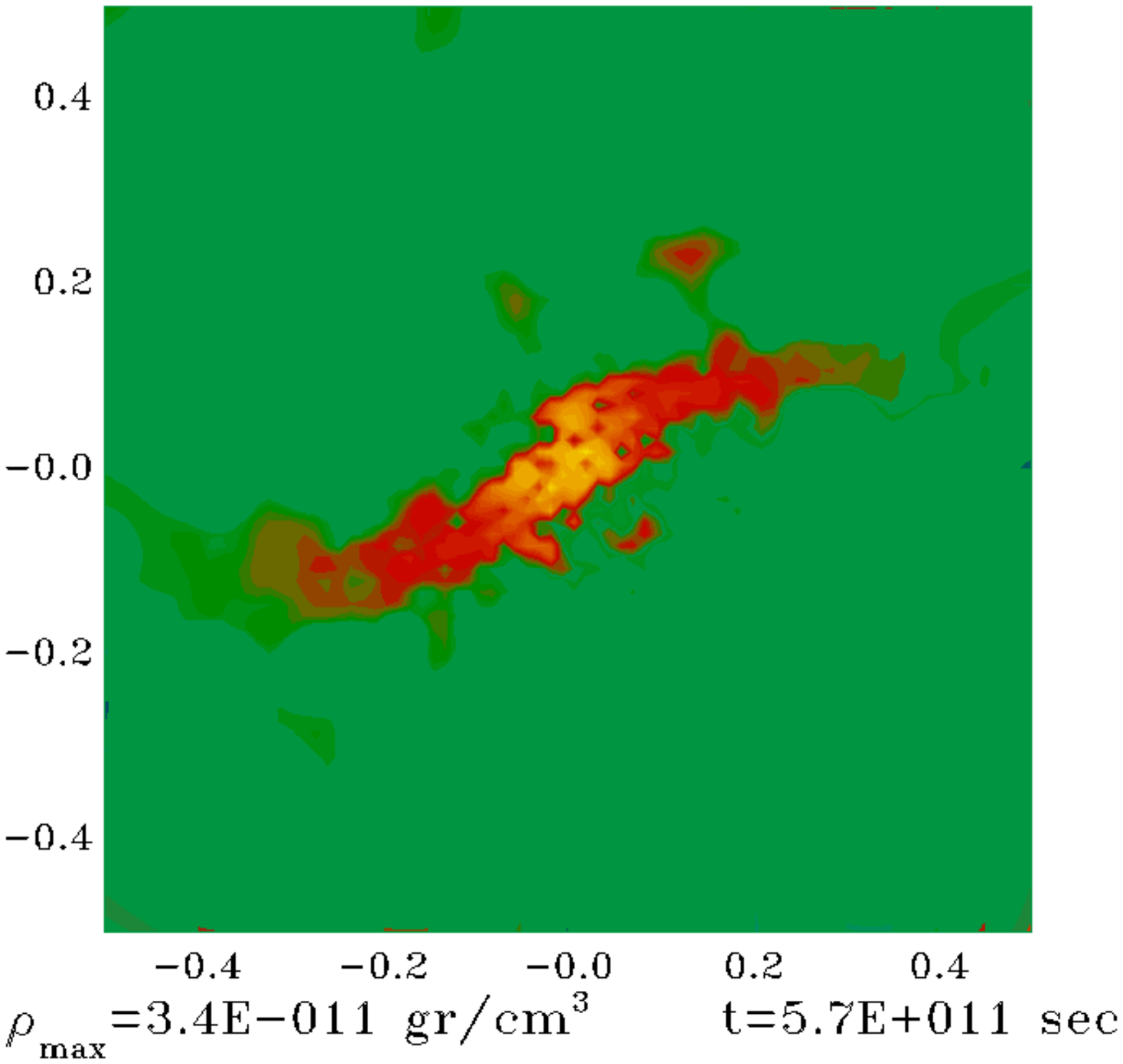}
& \includegraphics[width=2.2in]{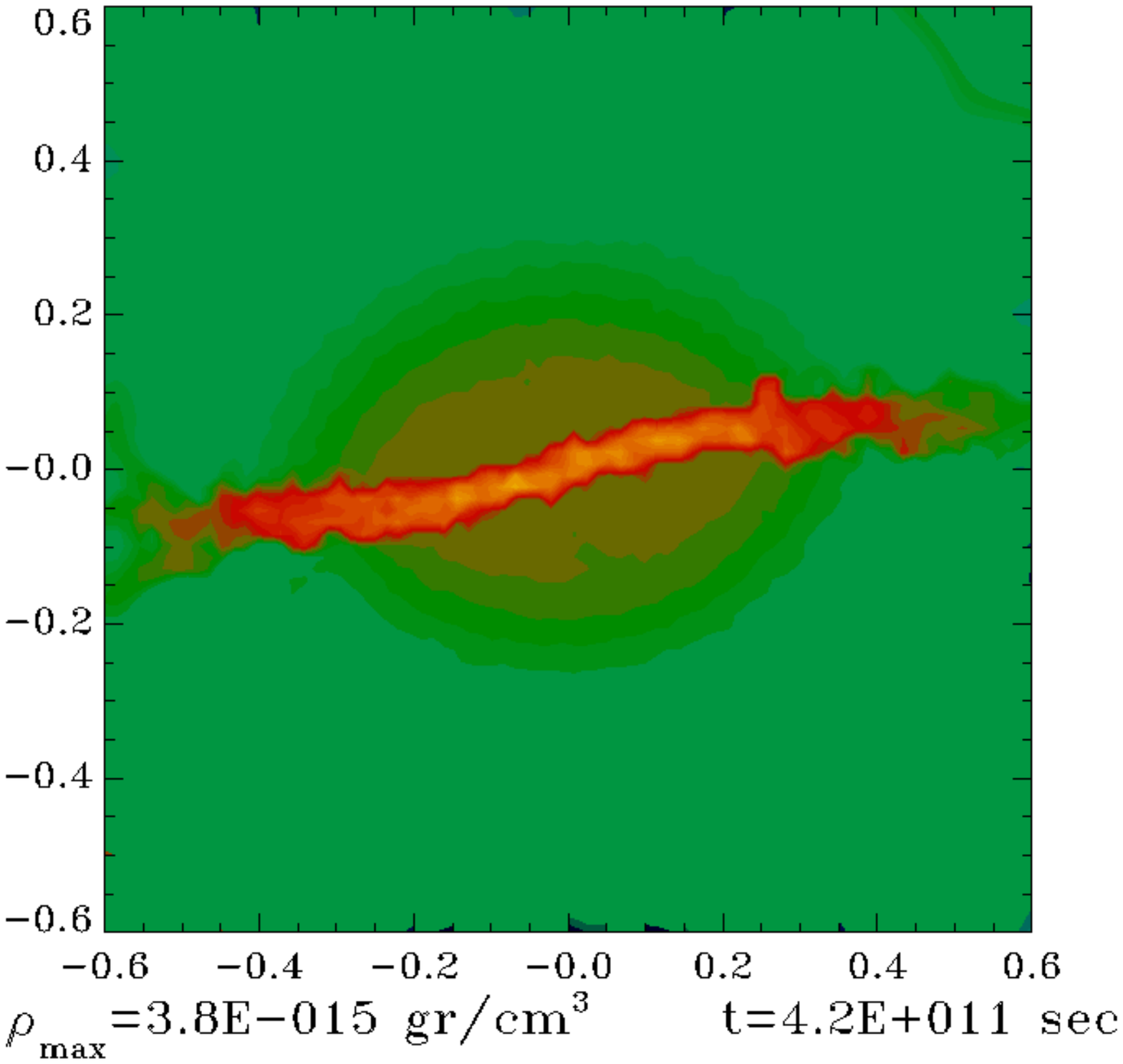}
\end{tabular}
\caption{ \label{ColH2} Isodensity plots for the last available snapshot in the
head-on collision models $HO-1$ (top left panel),
$HO-2$ (top middle panel), $HO-3$ (top right panel), $HO-6$ (bottom
left panel), $HO-5$ (bottom middle panel), and  $HO-7$ (bottom right
panel). } 
\end{center}
\end{figure}

In Fig. \ref{ColH2} we show the last snapshot available for each
collision model, or in other words, for the sake of brevity, we
omit presenting the earlier states of the evolution, like the first
three panels shown in Fig.~\ref{HO1norgad}.

With $V_{app}$ slightly less than $c_0$, which corresponds to model
$HO-3$, the initial cores are slightly disturbed, as the central
filament shows a little stretching.

Most of the gas in the filament remains there as the collapse
of each core progresses further. As we see in Fig.~\ref{modMCmR}, many accretion
centers are expected to be formed along these filaments, giving place to
the formation of a peculiar chain of protostars when the gas cools further.

\begin{figure}
\begin{center}
\includegraphics[width=2.2in]{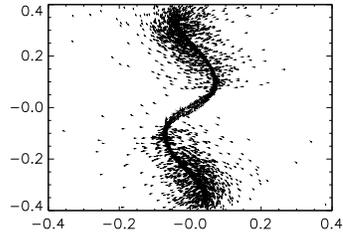} \\
\includegraphics[width=2.2in]{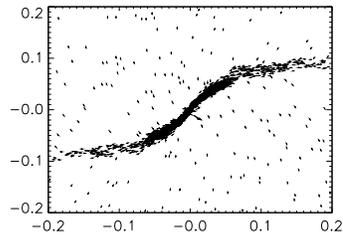}
\caption{ \label{VelHO2SHO6} Velocity field of the filament in models $H02$ 
(top) and $HO6$ (bottom), see the top middle panel and bottom left panel
of Fig. \ref{ColH2}, respectively.}
\end{center}
\end{figure}

In the top panel of Fig. \ref{VelHO2SHO6} we show the velocity field
of the particles located near the filament. It can be seen that
these particles flow into the filament. In Fig.~\ref{LocColM2p}, we
show a $3D$ visualization of the same filament of model $HO2$ seen
from a different angle; in this view, it is possible to observe that
a very small over-density is formed at each filament's end, at the
colliding interface between the two collapsing cores. The origin of
these new over-densities is the density perturbation caused by the
core collision.

With a higher pre-collision velocity, self-gravity will have less
time to act on the core and many more particles from the cores would
flow very quickly to the cores' interface, up to the point that the
two initial cores can no longer be distinguished, as can be seen
in models $HO-6$ and $HO-7$. For these models, the net effect of the
collision is then to replace the two initial colliding cores by only
one core, which rapidly forms a long central filament-like
structure. In the bottom panel of Fig. \ref{VelHO2SHO6}, we observe
again the formation of a strong filament in the interface of the
cores that expands rapidly as a strong flow of particles go by the
ends along the $X$-axis. However, the expansion eventually stops and
the filament bounces to re-initiate the expansion in the transversal
direction, along the $Y$-axis. This behavior has previously been
observed by \cite{pindika}.

Thus, one effect of having a higher pre-collision velocity is that the
density perturbation is stronger as can be seen in the last panel of Fig.
\ref{ColH2}.  If $V_{app}$ is high enough, we see that the net effect
of the collision is to replace the two initial colliding cores by
one core which may not be collapsing anymore.

\subsection{Oblique collisions}
\label{subsec:conbrazo}

\begin{figure}
\begin{center}
\begin{tabular}{ccc}
\includegraphics[width=2.2in]{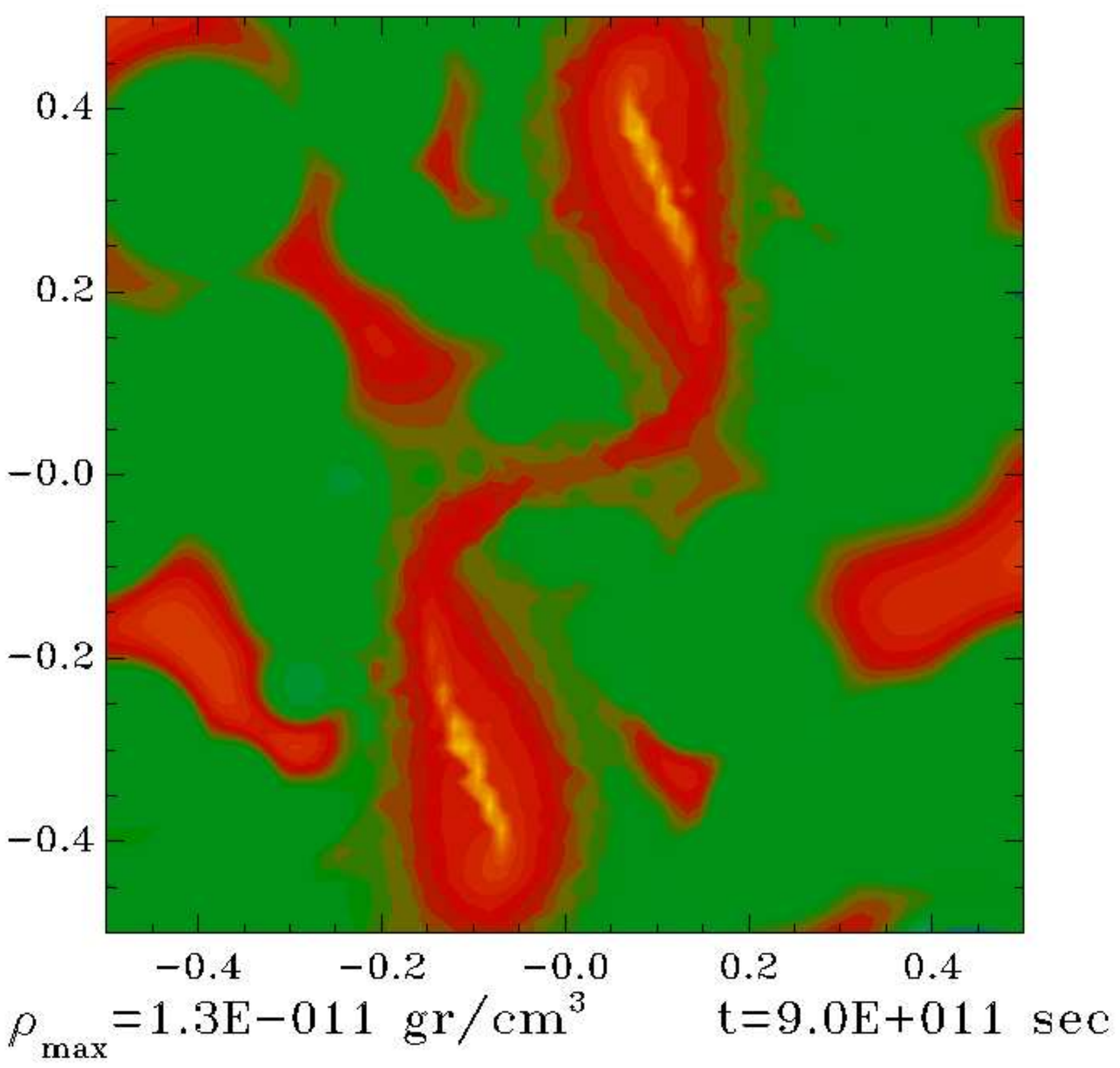} & 
\includegraphics[width=2.2in]{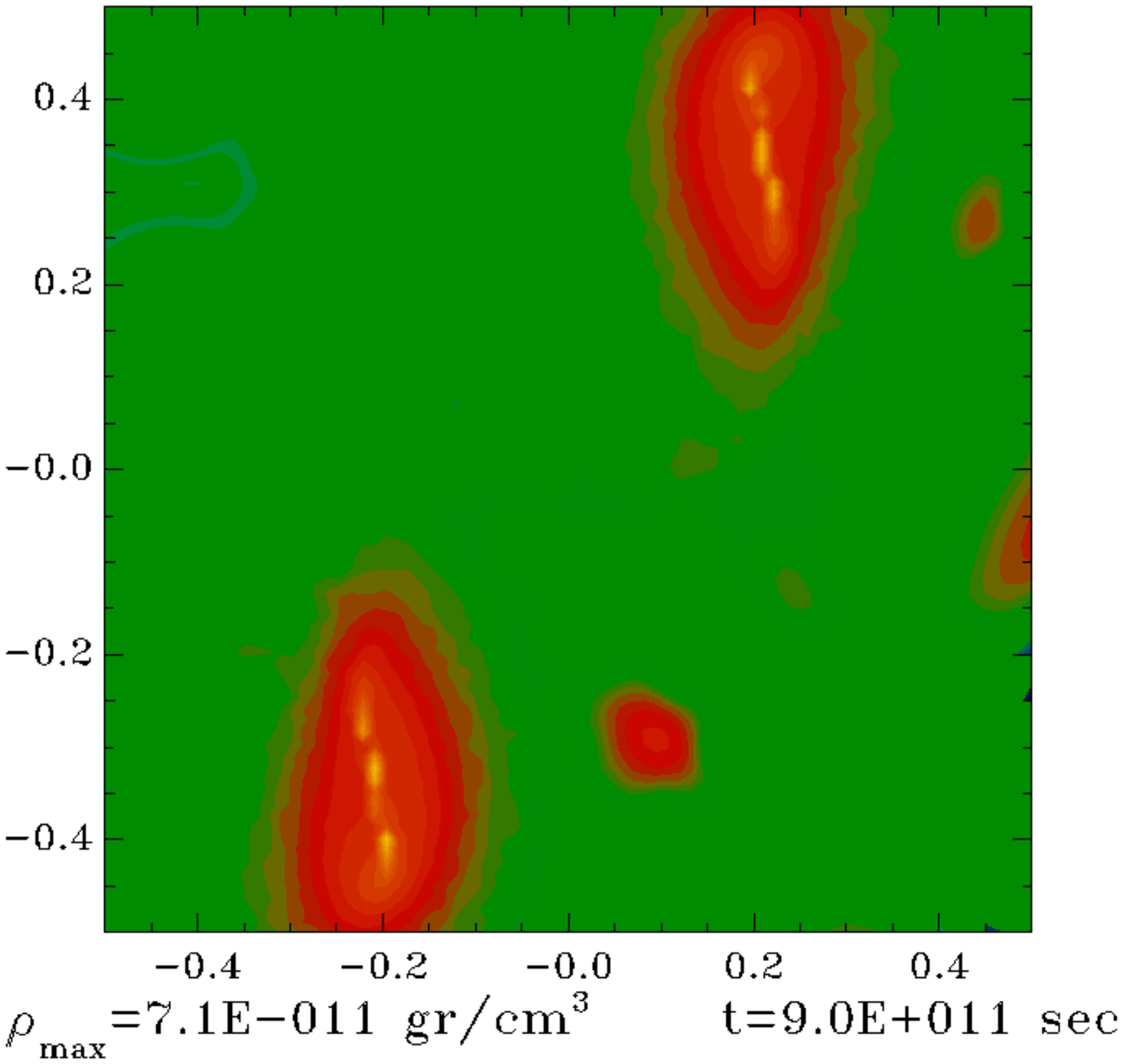} &
\includegraphics[width=2.2in]{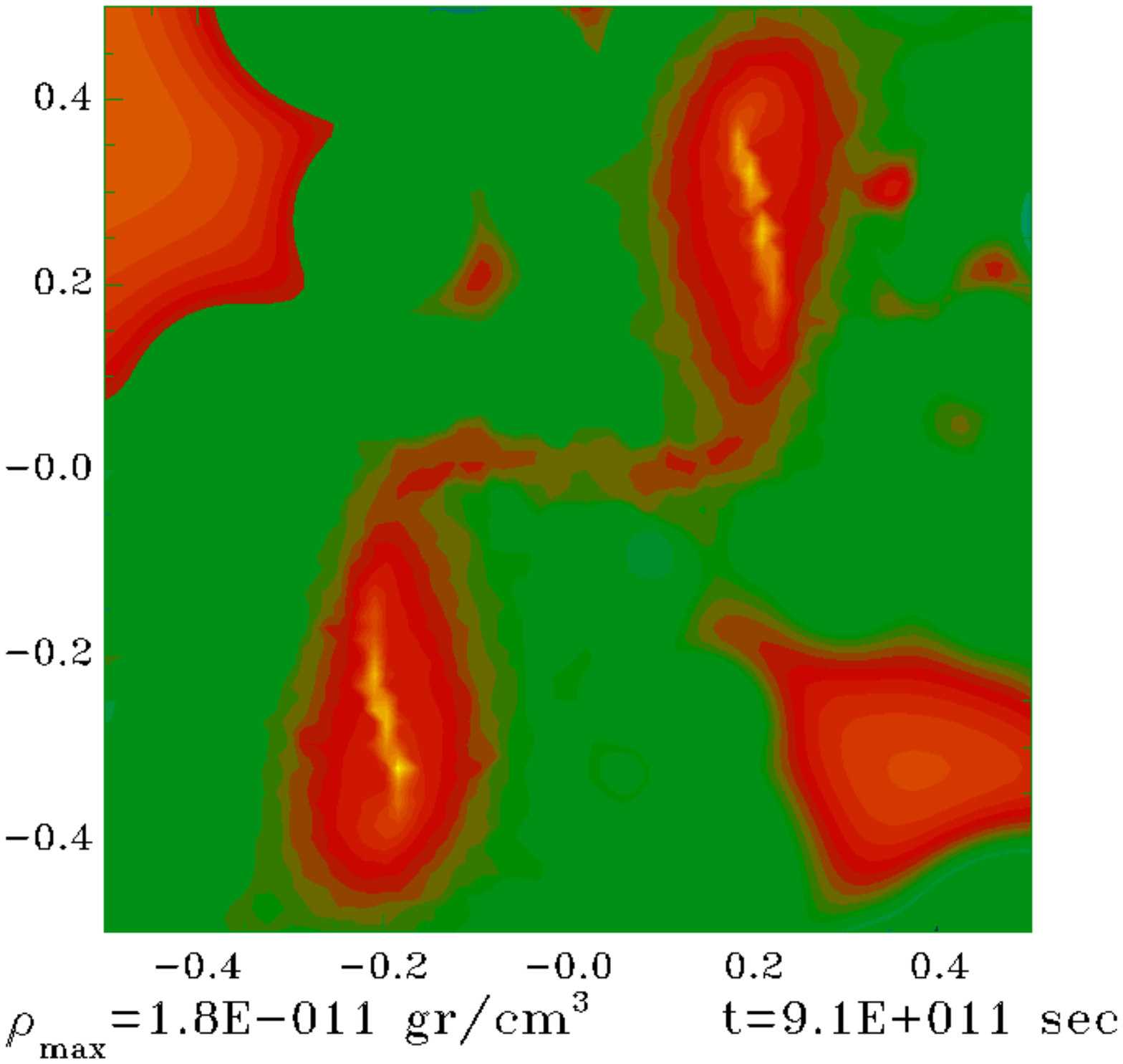} \\
\includegraphics[width=2.2in]{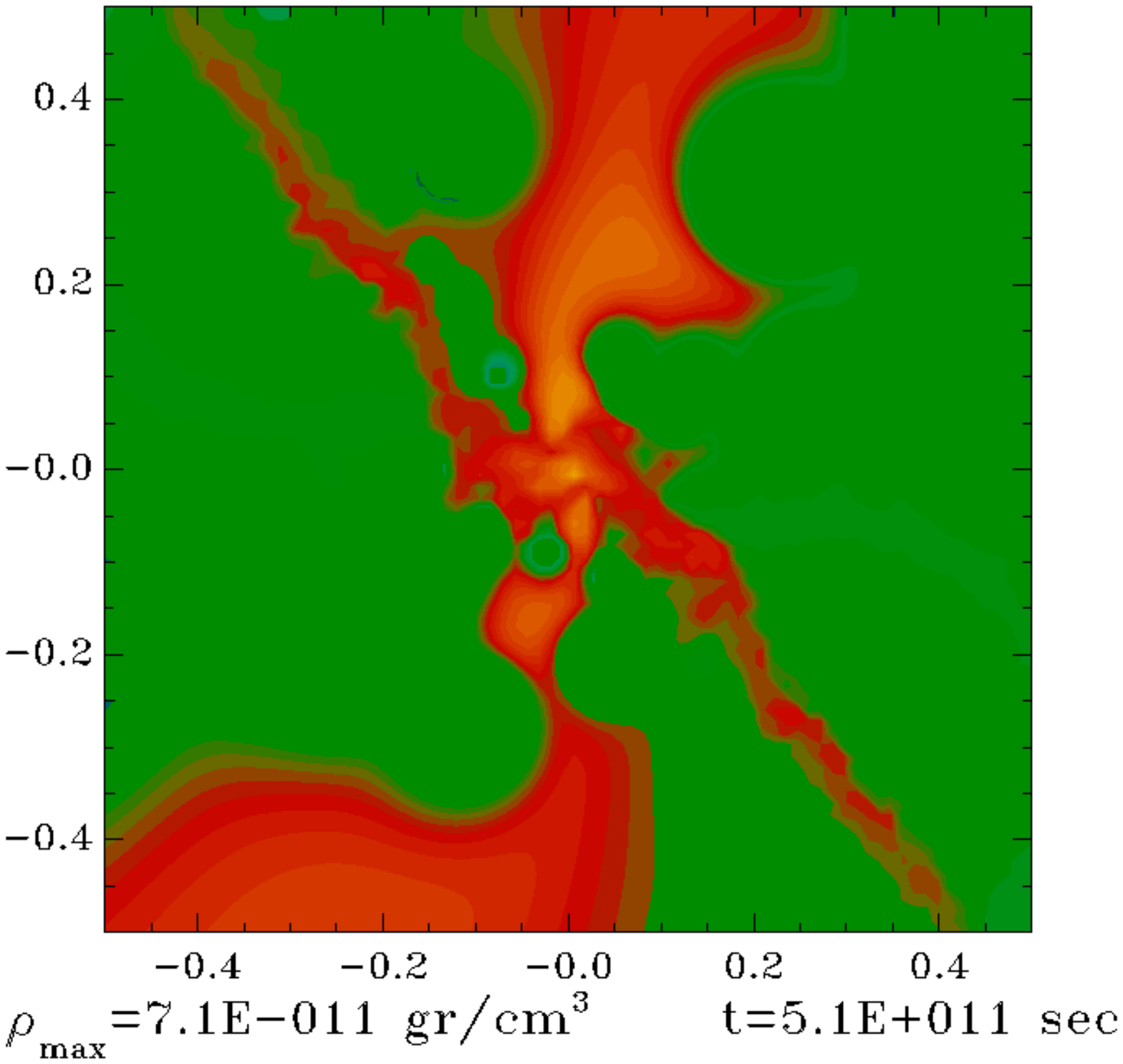} &
\includegraphics[width=2.2in]{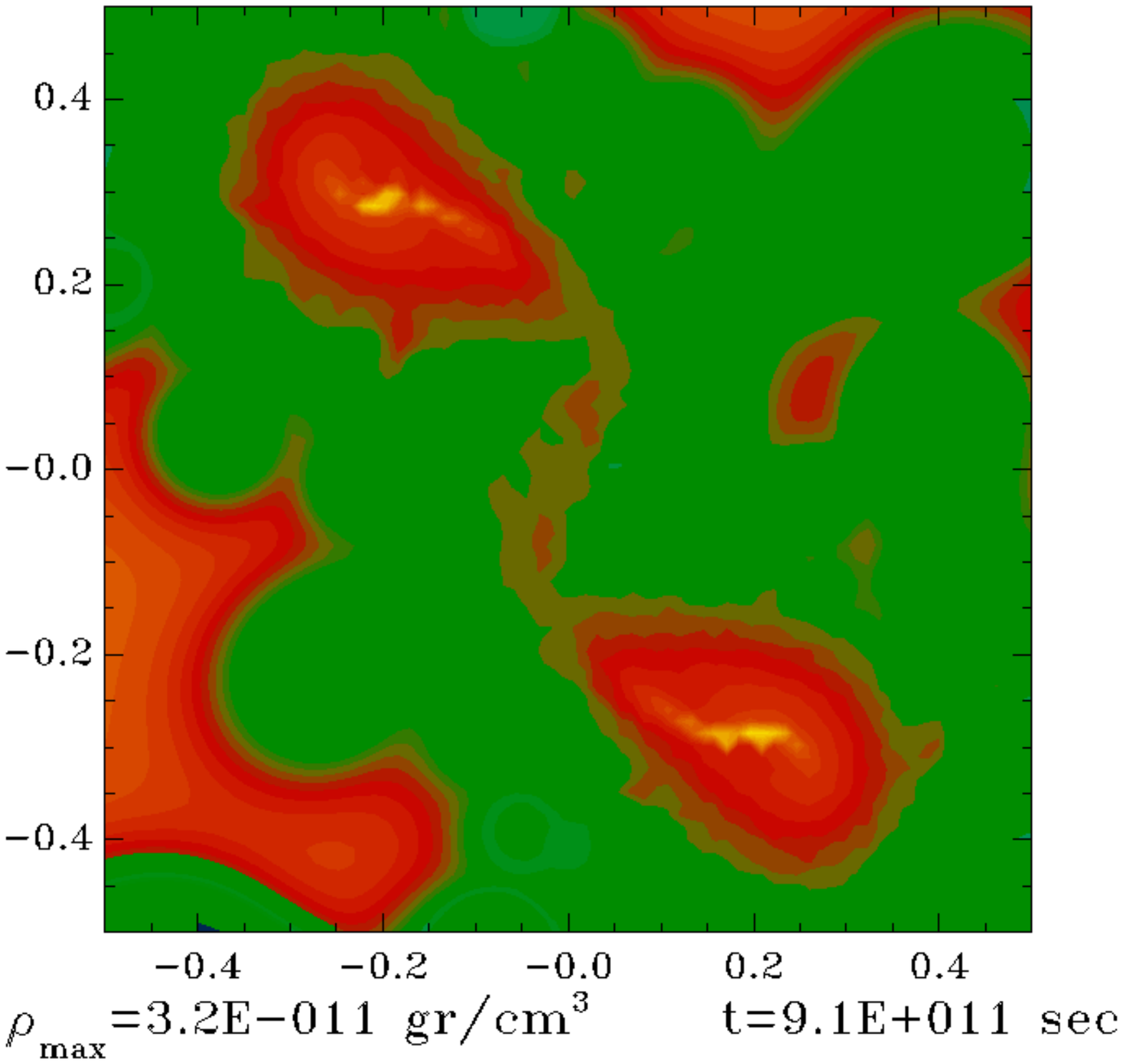} & 
\includegraphics[width=2.2in]{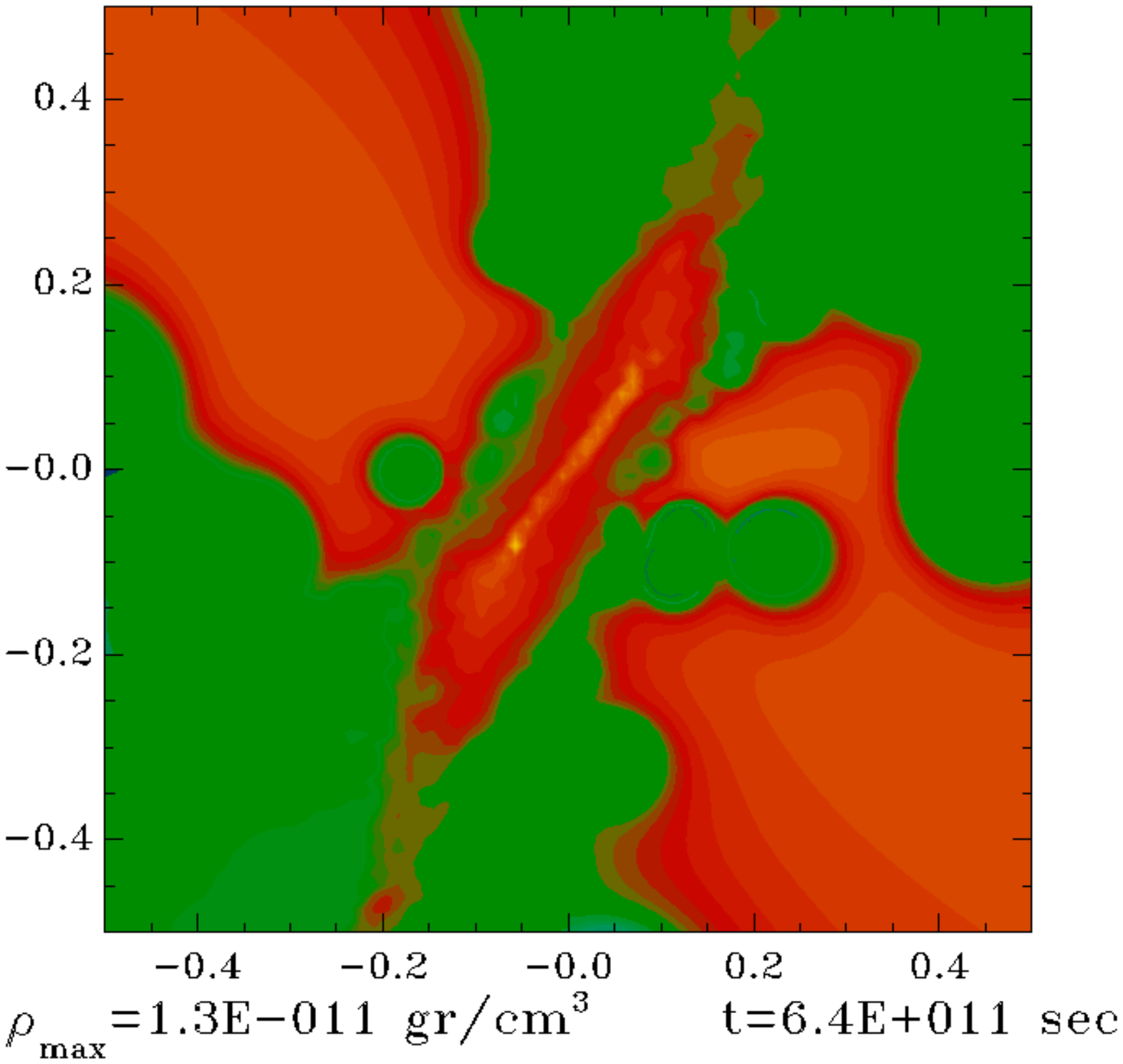} \\
\end{tabular}
\caption{\label{ColM2} Iso-density plots for the last available snapshot for the
oblique collision models $M1$ (top left panel), $M2$ (top middle
panel), $M21$ (top right panel), $M22$ (bottom left panel), $M3$
(bottom middle panel), and $M4$ (bottom right panel). }
\end{center}
\end{figure}

The outcome for the oblique models is illustrated in Fig.
\ref{ColM2}, where we show only the last available snapshot of each
model. As expected, the results for model $M1$, with a low
pre-collision velocity and with small impact parameter value, are
quite similar to those obtained for the head-on collision models
$HO-1$, $HO-2$ and $HO-3$. Hence, it is unnecessary to calculate more
models with increasing $V_{app}$ with a small impact parameter,
because we would expect to see similar results to those obtained for
the head-on collision models.

We then go on to models $M2$, $M21$ and $M22$, where the impact
parameter has been increased to $b=R_0/2$, again with the positive
orientation. The pre-collision speed is low for the two former models
and higher for the later model. Similar results are obtained for
models $M3$ and $M4$, where we change the orientation of the
impact parameter.

The net effect of a high $V_{app}$
on the colliding system is evident in models $M22$ and $M4$, as the
two cores get dispersed to give place to a long filament. It must be
emphasized that the orientation and the width of these
filaments are different from the ones obtained for
the high $V_{app}$ head-on collision models.
\subsection{Formation and location of the accretion centers}
\label{subsec:locacccent}

\begin{figure}
\begin{center}
\begin{tabular}{cc}
\includegraphics[width=2.2in]{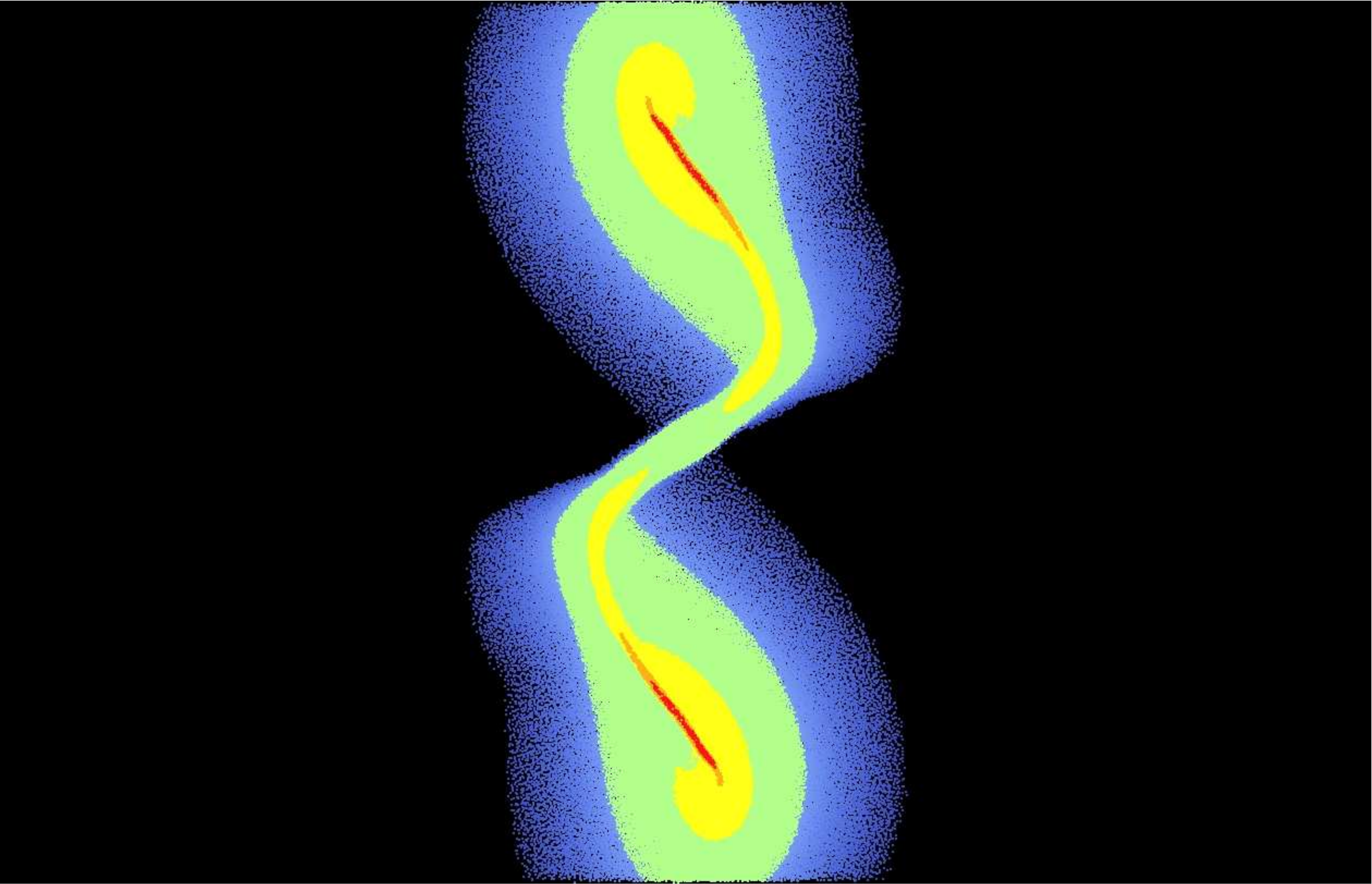} & 
\includegraphics[width=2.2in]{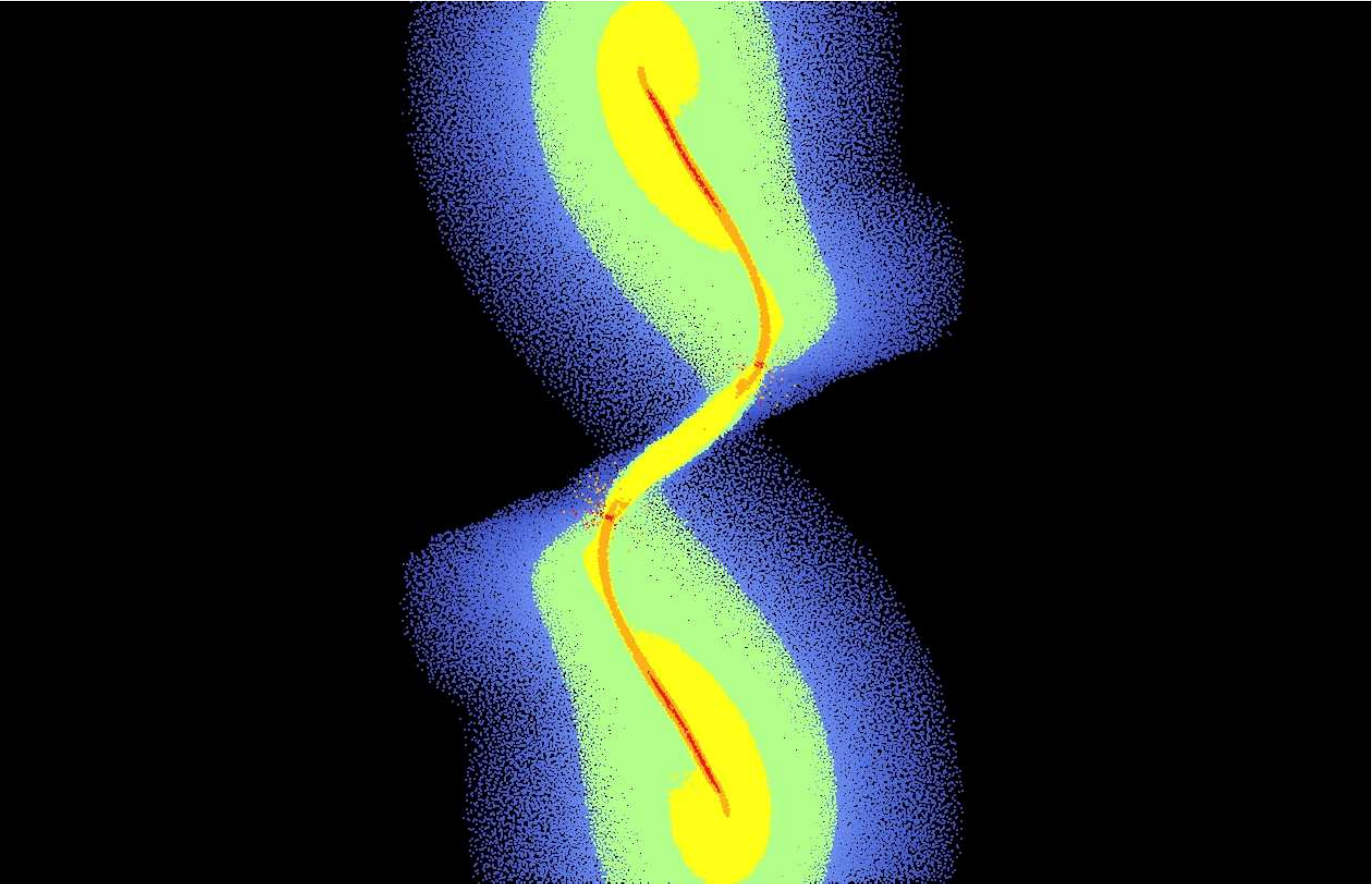} \\
\includegraphics[width=2.2in]{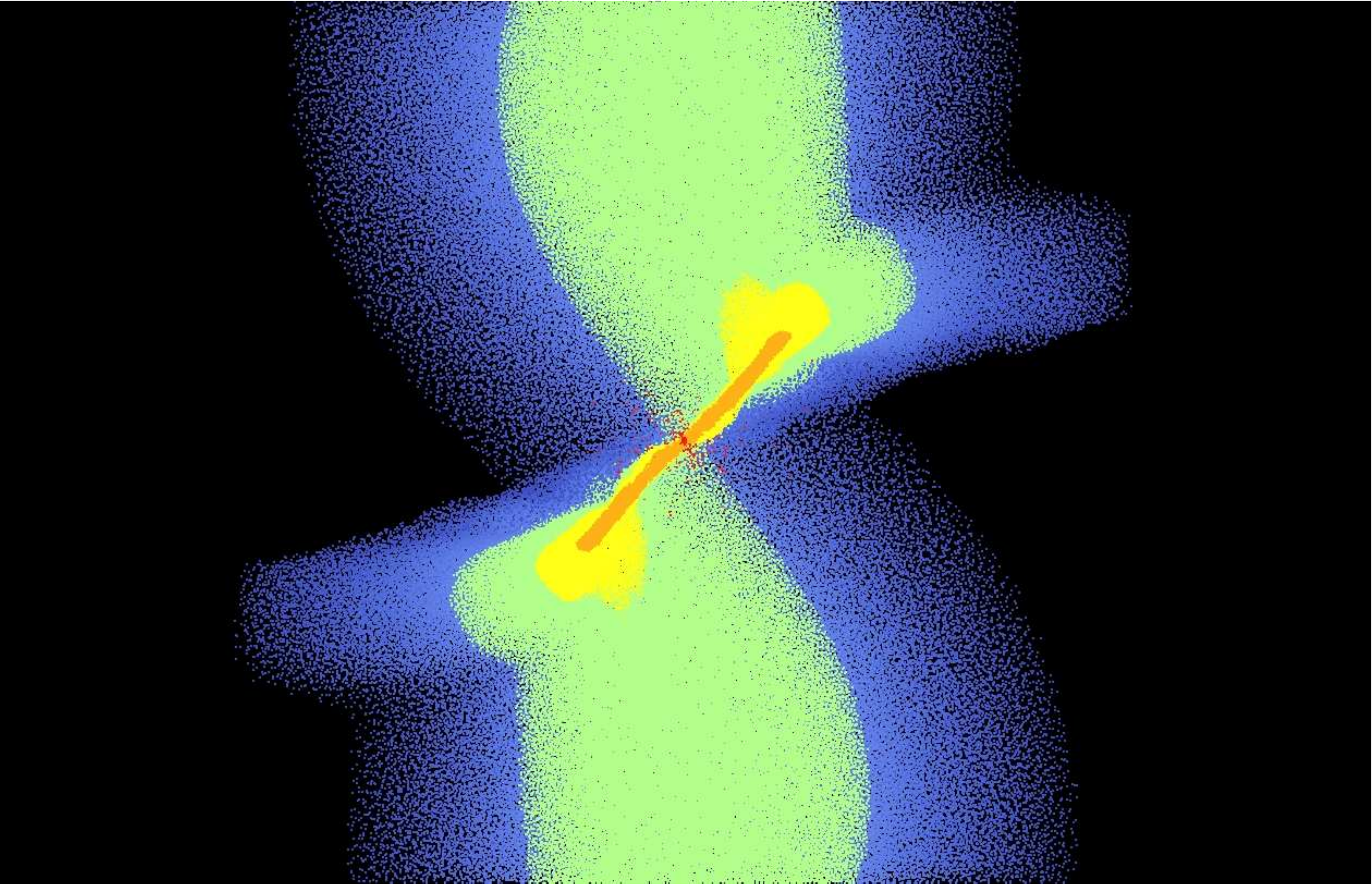} & 
\includegraphics[width=2.2in]{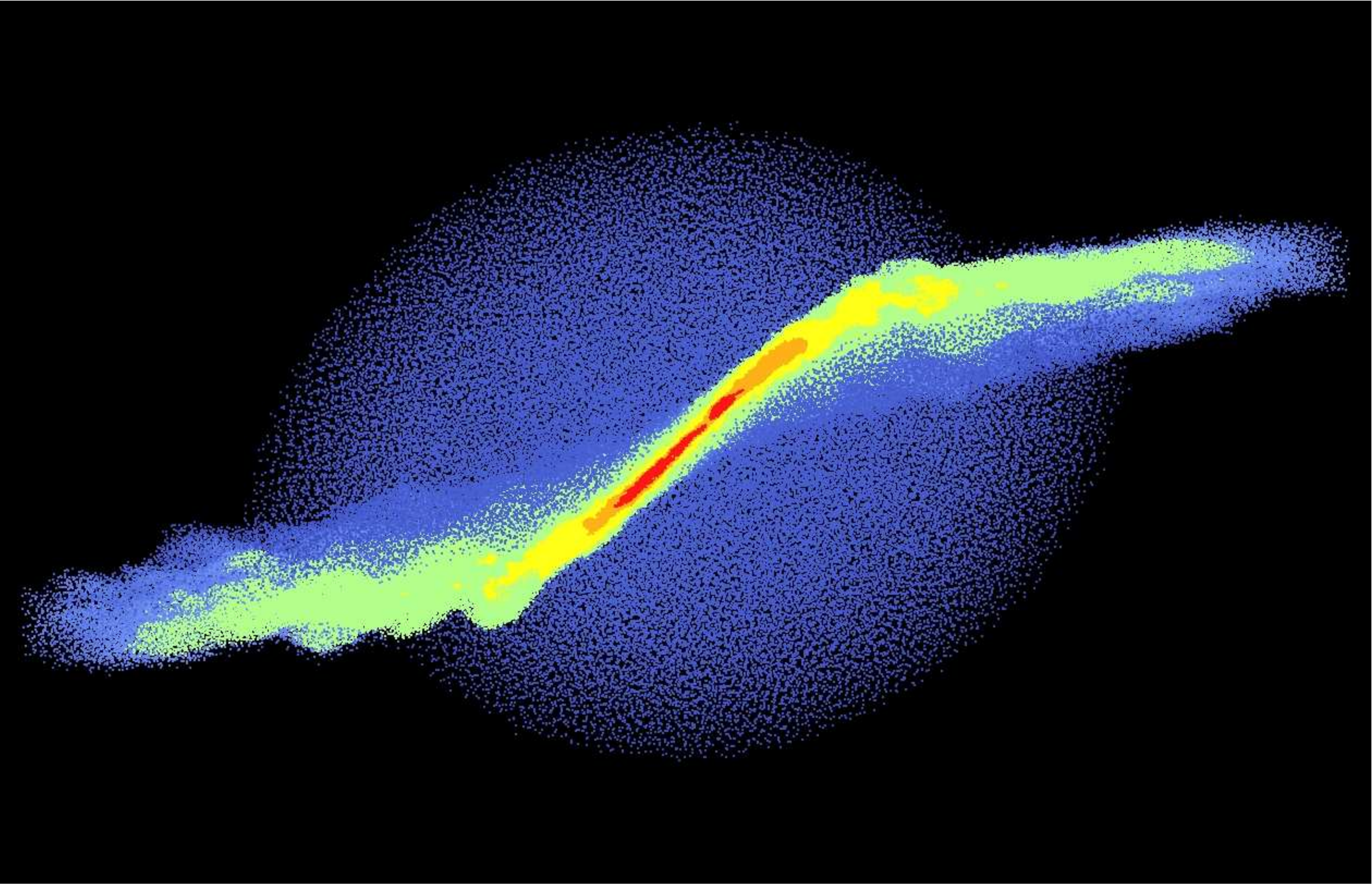} \\
\includegraphics[width=2.2in]{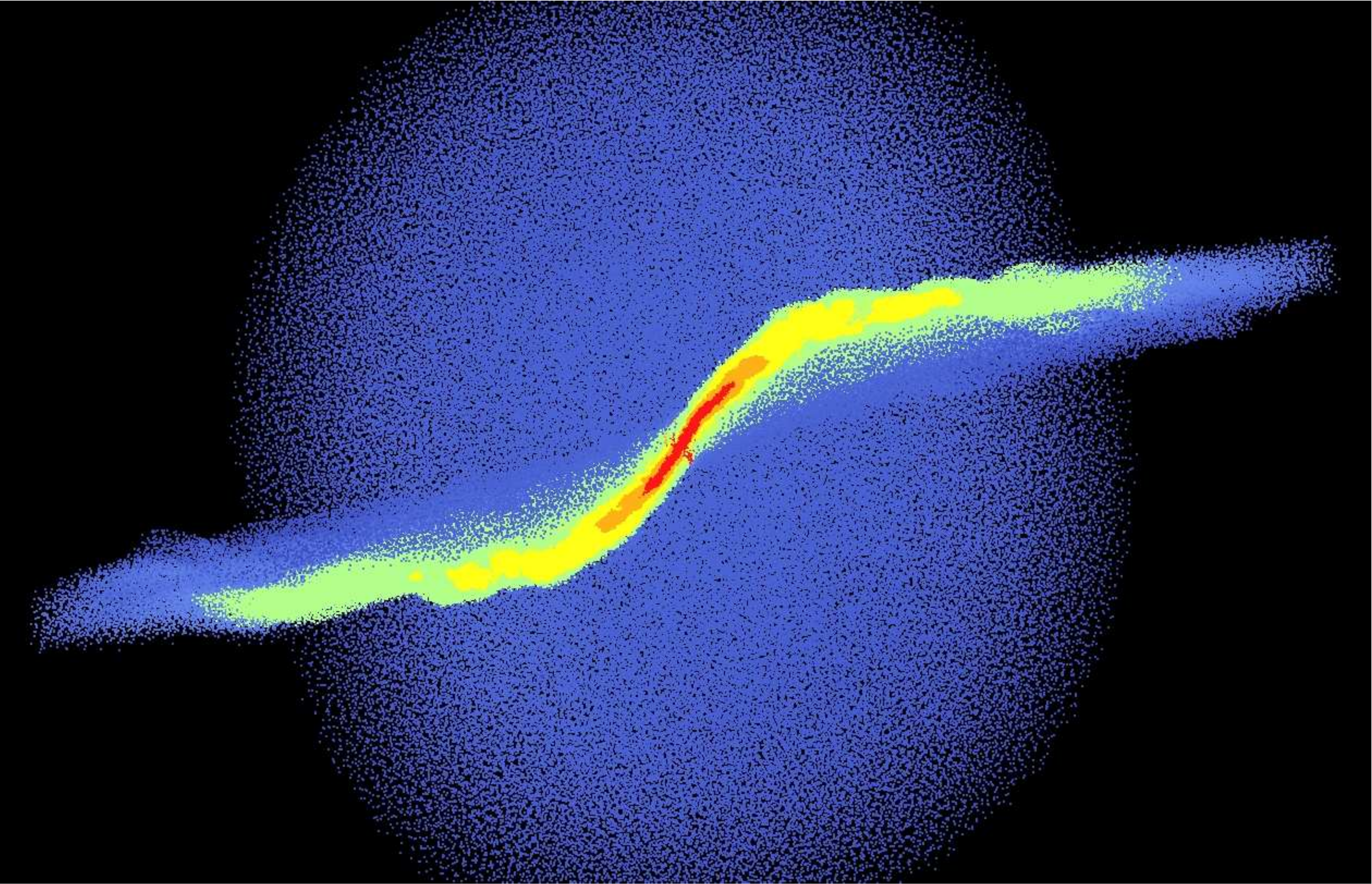} & 
\includegraphics[width=2.2in]{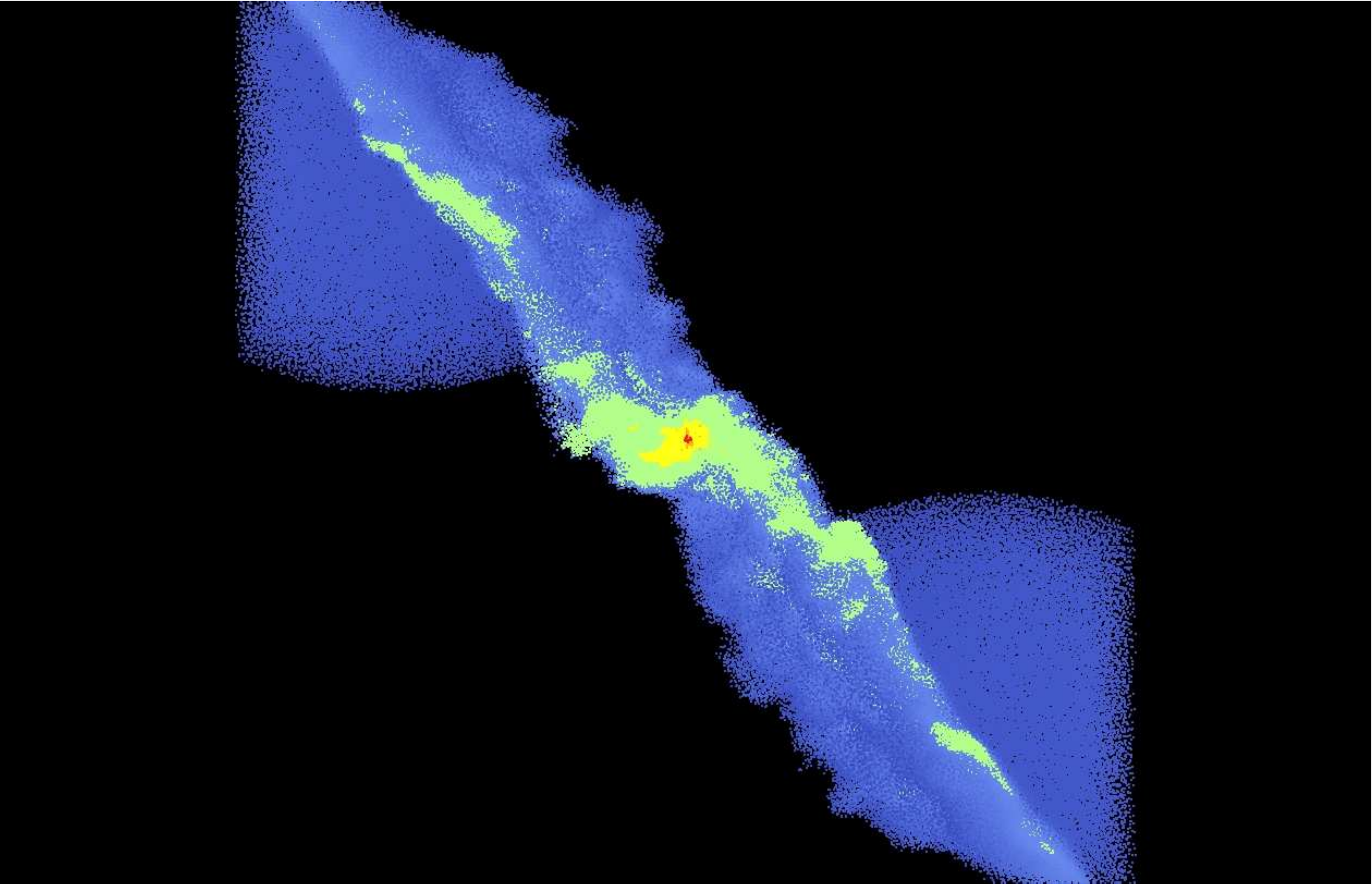} \\
\includegraphics[width=2.2in]{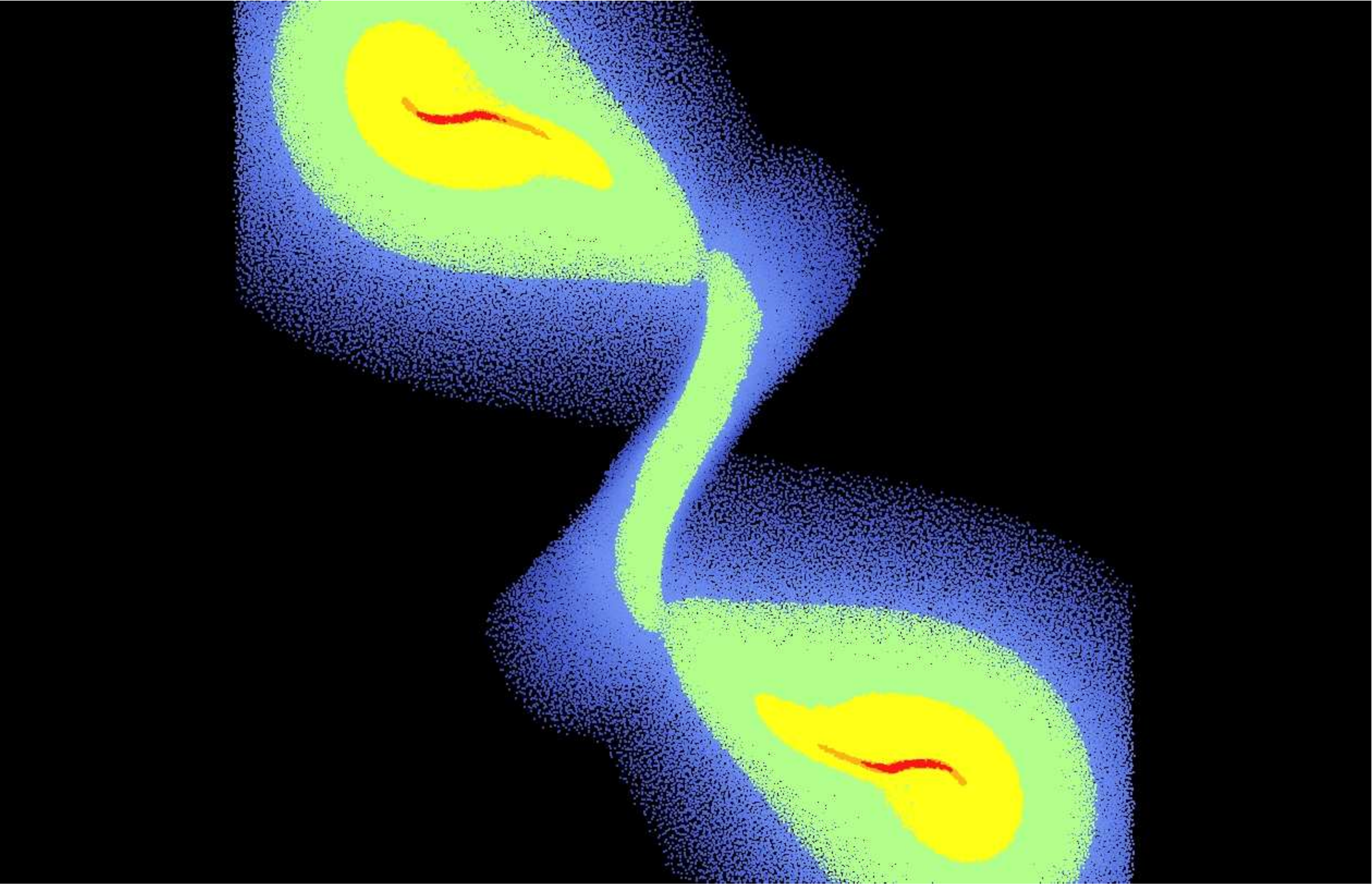} & 
\includegraphics[width=2.2in]{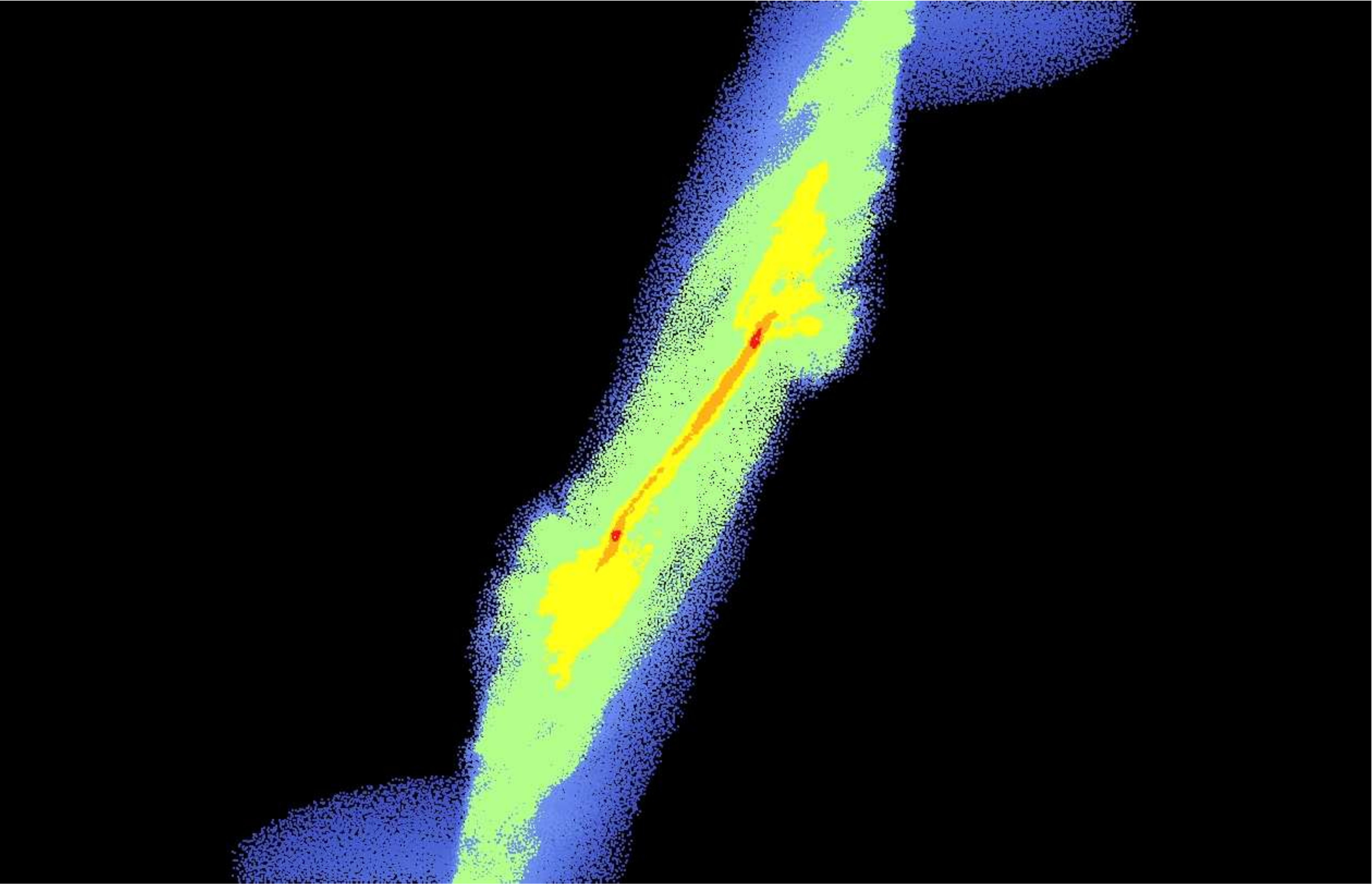}
\end{tabular}
\caption{ \label{LocColM2p} Location of the accretion centers for the 
last available snapshot
of models
$HO1$ (first line left), $HO2$ (first line right),
$HO3$ (second line left), $HO5$ (second line right),
$HO6$ (third line left), $M22$ (third line right),
$M3$ (fourth line left), and  $M4$ (fourth line right). The colors are a 
density scale
for $ \log_{10} \left( \frac{\rho}{\rho_0} \right) $ according with
the following ranges: blue ( $<2$), green ($2-3$), yellow ($3-4$), 
orange ($4-5$) and red ($>5$).  }
\end{center}
\end{figure}

In order to show the results of this section, we present $3D$ plots
of all the particles located in a central cube (with half the length
of the side of the original simulation box) and whose density is greater
than $1.2 \times 10^{-17} \,$ g $ \,$cm$^{-3}$, see Fig. \ref{LocColM2p}. As
in Section \ref{sec:resultados}, we again set a color scale
related to density: red and orange indicate particles
with higher densities, blue those with lower densities, and
and yellow the intermediate density particles, see the caption
of Fig.~\ref{LocColM2p}. It
should be noted that there is no relation between the density colors
associated with different panels even in the same plot. We rotate
each panel as needed for the purpose of better appreciating the
location of the accretion centers.

As expected, the pre-collision velocity $V_{app}$ also has a strong
influence on the number and location of accretion centers, as it
seems that the greater the $V_{app}$ the lower the $N_{acc}$. As
can be seen in Fig. \ref{LocColM2p}, if $V_{app}$ is too low, the
initial binary core system survives even to an advanced
stage of the evolution, and the accretion centers are mainly localized in
the core's central filament. But if  $V_{app}$ is too high, then the
initial binary core system is destroyed by the strong flow of
particles, giving less chance for accretion center to form. For
models $HO-3$ and $HO-6$, the accretion centers are indeed located
mainly in the cores' interface, as can be seen in the
close-up shown in Fig. \ref{LocColM2p}.

\begin{figure}
\begin{center}
\includegraphics[width=3.0in]{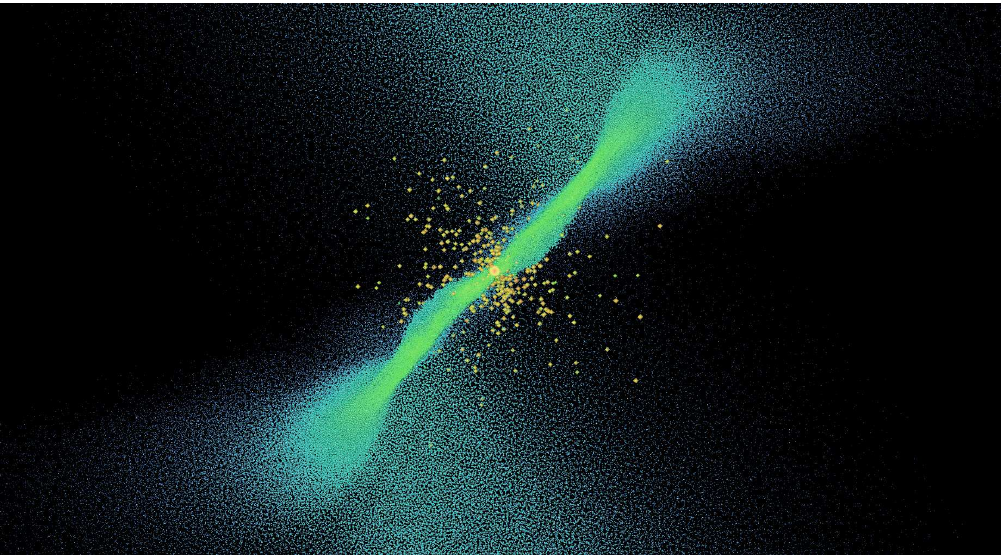}
\caption{\label{Vis3DHO3S} Zoom in of the left panel of the second 
line in Fig.~\ref{LocColM2p}, where the
accretion centers are shown for model $HO3$.}
\end{center}
\end{figure}

It is therefore remarkable to realize that models $HO-2$ and $HO-3$
have accretion centers forming outside the core's central
filament, as can be seen in Fig. \ref{LocColM2p}. As we can see, there
must be a small range of velocities $V_{app}$ which favors the formation of
accretion centers in the gas bridge connecting the two cores, see 
Fig.~\ref{Vis3DHO3S}.

From Table \ref{tab:modelos},  we first notice that the oblique
collision models have a larger number of accretion centers than the
head-on models.  It seems that more accretion centers are created if
self-gravity has more time available to act on the collapsing
cores. 

\section{Discussion}
\label{sec:dis}

For this paper we carried out a fully $3D$ set of numerical
hydrodynamical simulations within the framework of the SPH
technique, aimed at following the formation of accretion centers in a
collision process of two small rotating cores with uniform density.

The initial conditions for the isolated core are such that it will
collapse because its initial $(\alpha+\beta)_{core}<0.5$. However,
when we consider the translational kinetic energy that comes from
the pre-collision velocity $V_{app}$ of the cores, the collision
system is expected to be slightly unbound, that is,
$(\alpha+\beta)_{collision}>0.5$ for all models. As a result of
the collision process, a large fraction of the translational kinetic
energy will be transformed into heat, which will be radiated away
because the cores and the perturbation front that forms in the
interphase between the cores are isothermal. This must somehow
produce a reduction of the global value of $(\alpha+\beta)$ for our
collision models, as we observe that most of them are on the verge
of gravitational collapse, see Fig.~\ref{denmax}.

We have seen that the result of a collision simulation depends
mainly on two physical
factors that compete to exert influence on the collision
process. These factors are the self-gravity and the flow of the particles
from the colliding interface. The occurrence and extension of these
factors are regulated by the value of the
pre-collision velocity $V_{app}$ of the collision system.

With a very slow pre-collision velocity, self-gravity dominates the
evolution of both cores. In such cases, the main effect of
self-gravity on the system is to reduce the core size. The collision
system quickly reaches a configuration in an advanced state of
gravitational collapse, consisting of the remains of each collapsed
core linked by a bridge of matter. We note that the accretion centers in
the simulations are always located in the gas remnants of the cores.
In most collision cases, we expect these accretion centers to
separate from the filamentary structure, so that each fragment is
expected to evolve towards virial equilibrium.

On the other hand, when a collision takes place with a significant
pre-collision velocity, self-gravity plays a minor role and the fate
of the system is entirely determined by the strong flow of particles
in the interface of the cores. In these cases, the main effect on
the system is to replace the two initial collapsing cores by a
single core that rapidly gets perturbed to give place to an elongated
and irregular filamentary structure formed in the interaction region. This
filamentary structure increases its size very rapidly as it is
continuously fed by in-falling material coming from the original
cores. In these cases, we observe that accretion centers are
formed in the central region of this filament structure.

However, in view of the limited nature of our simulations, the
future of this filamentary structure is as yet unclear; but it is likely 
that most
of the gas already in the filament will stay in this filamentary
structure as the collapse progresses further. Although accretion
centers may be formed near the bridge edges, as we see in the case
of model $HO-2$, these accretion centers will be very thin and
small, so it could be the case that their diameter will be greater in
general than the Jeans length. This is another reason why
the bridge would not be an appropriate site for proto-stellar
formation until it cools further.

The results obtained for model $HO-3$ have been emphasized
by zooming in on the filament's central region, as shown
in Fig.~\ref{Vis3DHO3S}. We here see the formation of a
well defined central accretion center, which is surrounded by
many smaller accretion centers. The properties of this clustered
distribution of protostars will be discussed elsewhere.

Now, according to Table \ref{tab:tiempos}, the maximum evolution
time reached by the models is in the range $ 0.6 < t_{evol}/t_{ff} <
1.3$, where the free fall time $t_{ff}$ was defined in the first
paragraph of Section \ref{sec:nubeinicial}. Except for models
$HO-6$ and $HO-7$, all the other models have reached a $t_{evol}$
greater than $t_{ff}$. It should be noticed that most models have
therefore evolved in time up to an advanced stage of collapse.
Unfortunately, it is not possible to compute most of the models any
further, as the time step becomes very small.

The relevant time scale for the collision is given
approximately by $t_{col} \approx R_0 / V_{app} $. For
the models of this paper, this collision time is in the range $ 0.5
< t_{col}/t_{ff} < 12$. For model $HO-6$ these time scales are of
the same order, $t_{col}/t_{ff} \sim 1$. By looking at Fig.
\ref{ColH2}, we see that the pre-collision velocity of model $HO-6$
is the maximum collision velocity that allows an initial binary
colliding core system to remain as a binary system capable of
undergoing further gravitational collapse and eventually of
producing multiple protostars.  It is in this sense that
our results have to be compared with those of \cite{pongracic}, who found
that it is more likely that a colliding system will in
general result in disruption and dispersal of the core involved, with
no chance of forming protostars.

\section{Concluding Remarks}
\label{sec:conclu}

In this paper we have been able to demostrate some of the
essential features of the two core collision processes. In a first
approximation, we have investigated the number and location of the
accretion centers formed in this collision scheme.

The modified GADGET2 code tested here and described in Section
\ref{subs:code} is a first step towards the full sink-particle
technique implementation already in use worldwide. Our code is
still incomplete as we have not yet included all the particle tests
suggested by \cite{bate} and \cite{fede}. However, as we have
applied the same code to all the models, a comparison is
possible. The main results of this comparison are shown in Table
\ref{tab:modelos} and in Fig.~\ref{LocColM2p}. 

It must be taken into account that (i) we have used a low density
threshold $\rho_s$ for a particle to become an accretion center and
(ii) there is no restriction on the minimum number of particles that
must be captured for a particle to become an accretion center: so we
have seen that there are a large number of accretion centers having
just very few captured particles. For these two reasons we must
expect an overproduction of accretion centers in our simulations.
Finally, we recall that accretion centers are gas particles and can
collide or merge. In view of this, the future of the accretion
centers is still unclear, and requires taking the calculations
further in time.

Despite this, it is interesting to note that not all the
accretion centers in the simulations were always found located in
the remnants of the cores. There were also accretion centers found
in the bridge of matter, as was the case for model $HO-2$; and in
the central collision region, as occurred for models $HO-3$ and
$HO-6$. Hopefully all these accretion centers will end up as
proto-stellar seeds forming either a peculiar chained structure
or a clustered structure, respectively. For these reasons,
our hope is that the results presented here can be considered as further
evidence that core collisions may have an important influence on the
star formation process.

\section{acknowledgements}
  
We would like to thank ACARUS-UNISON, the Instituto Nacional de
Investigaciones Nucleares and the Cinvestav-Abacus for the use of
their computing facilities. The authors are grateful for financial
support provided by the Consejo Nacional de Ciencia y
Tecnolog\'{\i}a (CONACyT) - EDOMEX-2011-C01-165873.

\newpage
\newpage
\begin{table}[h]
\begin{center}
\caption{\label{tab:modelos} The collision models. }
\begin{tabular}{|c|c|c|c|c|c|c|}
\hline
{Model} & $\frac{b}{R_0}$ & $\frac{V_{app}}{c_0}$ & $ N_{acc} $ & 
$Np_{acc}$ & $ M_{av}/M_{\odot}$ & $\rho_{max}\; [gr/cm^3]$ \\
\hline
  MCmR  & -- & --  & 27 & 52 & $5.80\times 10^{-5}$ & $1.17 \times 10^{-10}$ \\
\hline
  HO-1  & 0  &0.23 & 80 & 22 & $1.97\times 10^{-5}$ & $7.05 \times 10^{-12}$ \\
\hline
  HO-2  & 0  &0.46 & 29 & 5  & $4.20\times 10^{-6}$ & $6.7 \times 10^{-11}$\\
\hline
  HO-3  & 0  &0.92 & 30 & 6 & $5.0\times 10^{-6}$ & $ 2.8 \times 10^{-10}$\\
\hline
  HO-5  & 0  & 3.9 & 104& 5 & $ 3.9 \times 10^{-6}$ & $ 3.37 \times 10^{-11}$\\
\hline
  HO-6  & 0  & 2.9 & 37 & 5 & $4.0 \times 10^{-6}$ & $ 1.0 \times 10^{-10}$\\
\hline
  HO-7  & 0  & 5.7 & 9 & 57 & $4.6 \times 10^{-5}$ & $ 1.66 \times 10^{-13}$\\
\hline
  M1    & 1/4 & 0.23 & 210 & 7 & $5.9 \times 10^{-6}$ & $ 1.2 \times 10^{-11}$\\
\hline
  M2    & 1/2 & 0.25 & 103 & 15 & $1.2 \times 10^{-5}$ & $  7.0 \times 10^{-12}$\\
\hline
  M22   & 1/2 & 6.4  & 42  & 5 & $ 4.4 \times 10^{-6}$ & $7.1 \times 10^{-11}$\\
\hline
  M21   & 1/2 & 0.51 & 214 & 4 & $4.2 \times 10^{-6}$ & $ 1.8 \times 10^{-11}$\\
\hline
  M3    &-1/2 & 0.51 & 156 & 5 & $4.3 \times 10^{-6}$ & $ 1.8 \times 10^{-11}$ \\
\hline
  M4    &-1/2 & 6.4  & 30  & 4  & $ 4.0\times 10^{-6}$ & $ 1.2 \times 10^{-11}$ \\
\hline
\end{tabular}
\end{center}
\end{table}
\begin{table}
\begin{center}
\caption{\label{tab:tiempos} The evolution and collision time of the collision models.}
\begin{tabular}{ccc}
\hline \\
{Model} &  $t_{evol}/t_{ff}$ & $t_{col}/t_{ff}$ \\
\hline
  HO-1  &  1.3 & 12.6 \\
\hline
  HO-2  &  1.3 & 6.3\\
\hline
  HO-3  &  1.1&  3.15\\
\hline
 HO-5   &  0.82 & 0.74 \\
\hline
  HO-6  &  0.9 & 1.0\\
\hline
  HO-7  &  0.75 & 0.5 \\
\hline
  M1    &  1.3 & 12.6 \\
\hline
  M2    &  1.3 & 12.6 \\
\hline
  M21   &  1.3 & 5.6 \\
\hline
  M22   &  0.74& 0.45   \\
\hline
  M3    &  1.3 & 5.6 \\
\hline
  M4    &  0.92& 0.45 \\
\hline
\end{tabular}
\end{center}
\end{table}

\newpage

\begin{thebibliography}{99}
  
\bibitem[Anathpindika (2009a)]{pindika} Anathpindika, S. 2009a,
A\&A, 504, 437.

\bibitem[Anathpindika (2009b)]{pindika2} Anathpindika, S. 2009b,
A\&A, 504, pp.451-460.

\bibitem[Anathpindika (2010)]{pindika3} Anathpindika, S. 2010,
MNRAS, 405,  pp.1431-1443.

\bibitem[Arreaga et al. (2007)]{NuestroApJ} Arreaga, G., Klapp, J., 
Sigalotti, L.G.,
and Gabbasov, R. 2007, ApJ, 666, 290.

\bibitem[Arreaga et al. (2008)]{NuestroRMAA}
Arreaga, G., Saucedo, J., Duarte, R., and Carmona, J. 2008, Rev. Mex.
Astron. Astrophys, 44, 259.

\bibitem[Arreaga \& Klapp (2010)]{NuestroPlummer} Arreaga, G. and Klapp, J. 
2010, A\&A, 509, A96

\bibitem[Bate \& Burkert (1997)]{bateburkert97} Bate, M.R. and Burkert, A., 1997,
MNRAS, 288, 1060.

\bibitem[Bate et al. (1995)]{bate}  Bate, M.R., Bonnell, I.A. and Price, M.N.,
1995, MNRAS, 277, pp.362-376.

\bibitem[Bergin \& Tafalla (2007)]{bergin} Bergin, E. and Tafalla, M. 2007, 
Annu. Rev. Astro. Astrophys.,
45, 339.

\bibitem[Balsara (1995)]{balsara1995} Balsara, D. 1995, J. Comput. Phys.,
121, 357.

\bibitem[Bekki et al. (2004)]{bekki} Bekki, K., Beasley, M., Forbes, D., 
and Couch, W.J. 2004,
ApJ, 602, 730.

\bibitem[Bodenheimer et al. (2000)]{bodeniv} Bodenheimer, P., Burkert,
A., Klein, R.I., \& Boss, A.P. 2000, in {\it Protostars and Planets
IV}, Eds. V.G. Mannings, A.P. Boss and S.S. Russell, University of
Arizona Press, Tucson.

\bibitem[Burkert \& Alves (2009)]{burkert} Burkert, A. and Alves,
J. 2009, ApJ, 695, 1308.

\bibitem[Boss et al. (2000)]{boss2000} Boss, A.P., Fisher, R.T.,
Klein, R. and McKee, C.F. 2000, ApJ, 528, 325.

\bibitem[Churchwell et al. (2006)]{churchwell} Churchwell, E., Povich, M.S., 
Allen, D., Taylor, M.G., Meade, M.R., Babler, B.L., Indebetouw, R., Watson, C., 
Whitney, B.A., Wolfire, M.G., Bania, T.M., Benjamin, R.A., Clemens, D.P., 
Cohen, M., Cyganowski, C.J., Jackson, J.M., Kobulnicky, H.A., Mathis, J.S., 
Mercer, E.P., Stolovy, S.R., Uzpen, B., Watson, D.F. and Wolff,M.J.,
2006, ApJ, 649, 759-778.

\bibitem[Clark \& Bonnell (2006)]{clark} Clark, P.C and Bonnell, I.A., 2006, 
MNRAS, 368, pp.1787-1795.

\bibitem[Duarte-Cabral et al. (2010)]{duarte} Duarte-Cabral, A., Fuller, G.A., 
Peretto, N., Hatchell, J., Ladd, E.F., Buckle, J., Richer, J and 
Graves, S.F., 2000, A\&A, 519, 27.

\bibitem[Federrath et al. (2010)]{fede} Federrath, C., Banerjee, R., Clark, P.C. and
Klessen, R.S., 2010, ApJ, 713, pp. 269-290.

\bibitem[Federrath \& Klessen (2012)]{fede2012} Federrath, C. and 
Klessen, R.S., 2012, ApJ, 761, pp. 156.

\bibitem[Furukawa et al. (2009)]{furu} Furukawa, N., Dawson, J.R., Ohama, A., 
Kawamura, A., Mizuno, T and Fukui, Y., 2009, ApJ, 696, pp. L115.

\bibitem[Gomez et al. (2007)]{gom} Gomez, G.C., Vazquez-Semanedi, E., 
Shadmehri, M and Ballesteros-Paredes, J., 2007, ApJ, 669, pp. 1042-1049.

\bibitem[Graves et al. (2010)]{graves} Graves, S. F.; Richer, J. S.; 
Buckle, J. V.; Duarte-Cabral, A.; Fuller, G. A.; Hogerheijde, M. R.; 
Owen, J. E.; Brunt, C.; Butner, H. M.; Cavanagh, B.; Chrysostomou, A.; 
Curtis, E. I.; Davis, C. J.; Etxaluze, M.; Francesco, J. Di; Friberg, P.; 
Friesen, R. K.; Greaves, J. S.; Hatchell, J.; Johnstone, D.; Matthews, B.; 
Matthews, H.; Matzner, C. D.; Nutter, D.; Rawlings, J. M. C.; Roberts, J. F.; 
Sadavoy, S.; Simpson, R. J.; Tothill, N. F. H.; Tsamis, Y. G.; Viti, S.; 
Ward-Thompson, D.; White, G. J.; Wouterloot, J. G. A.;
Yates, J. ,2010, MNRAS, 409, pp. 1412-1428.

\bibitem[Hausman (1978)]{hausman} Hausman, M.A. 1978, ApJ, 245, 72.

\bibitem[Heitsch et al. (2008)]{heitsch08} Heitsch, F., Hartmann, L.W.,
Slyz, A.D., Devriendt, J.E.G., and Burkert, A. 2008, ApJ, 674, 316.

\bibitem[Jijina et al.(1999)]{jijina} Jijina, J., Myers, P.C. and 
Adams, F.C., 1999,
ApJ Supplement Series, 125, 161-236.

\bibitem[Kimura \& Tosa (1996)]{kimura} Kimura, T. and Tosa,
M. 1996,  A\&A, 308, 979.

\bibitem[Kitsionas \& Whitworth (2007)]{kitsionas} Kitsionas, S. and 
Whitworth, A.P.,
2007,  MNRAS, 378, 507-524.

\bibitem[Klein \& Woods (1998)]{klein} Klein, R.I.and Woods,
D.T. 1998, ApJ, 497, 777.

\bibitem[Klessen \& Burkert (2000)]{klessen} Klessen, R.S. and 
Burkert, A., 2000, ApJS, 128, 
1, pp.287-319.

\bibitem[Larson (1981)]{larson81} Larson, R. 1981, MNRAS, 194, 809.

\bibitem[Lattanzio et al. (1985)]{lattanzio} Lattanzio, J.C., Monaghan, J.J.,
Pongracic, H. and Schwarz, M.P. 1985, MNRAS, 215, 125.

\bibitem[Marinho \& Lepine (2000)]{marinho} Marinho, E.P. and Lepine, J.R.D.
2000, A\&A Suppl., 142, 165.

\bibitem[Monaghan \& Gingold (1983)]{mona1983} Monaghan, J.J. and
Gingold, R.A. 1983, J. Comput. Phys., 52, 374.

\bibitem[Motte et al. (1998)]{motte} Motte, F., Andr\'e, P and Neri, R. 1998,
Astro.Astrophys, 336, 150-172.

\bibitem[Nelson \& Papaloizou (1994)]{papa} Nelson, R.P. and Papaloizou, C.B., 1994,
MNRAS, 270, pp.1-20.

\bibitem[Padoan et al.(2014)]{padoan} Padoan,P., Federrath, C., Chabrier, G., 
Evans, N.J. II, 
Johnstone, D., Jorgensen, J.K., McKee, C.F. and Nordlund, A., 2014, 
Protostars and Planets VI, Beuther, H., Klessen, R.S., Dullemond, C.P. and 
Henning, T., (eds). 
University of Arizona Press, Tucson, 914.pp.-pp.77-100.

\bibitem[Sigalotti \& Klapp (2001a)]{sigalotti2001a} Sigalotti, L.G. and 
Klapp, J. 2001,
International Journal of Modern Physics D, 10, 115.

\bibitem[Sigalotti \& Klapp (2001b)]{sigalotti2001b} Sigalotti, L.G. and 
Klapp, J. 2001,
A\& A, 378, 165-179.

\bibitem[Scoville et al. (1986)]{scoville} Scoville, N.Z., Sanders,
D.B. and Clemens, D.P. 1986, ApJ, 310, L77.

\bibitem[Springel (2005)]{gadget2} Springel, V. 2005, MNRAS, 364, 1105.

\bibitem[Tackenberg et al.(2014)]{tackenberg} Tackenberg, J. , Beuther, H.,
Henning, T. , Linz, H., Sakai, T., Ragan, S.E., Krause, O., Nielbock, M.
Hennemann, M., Pitann, J. and  Schmiedeke1, A.,2014, arXiv.1402.0021,
accepted by A\& A.

\bibitem[Tafalla et al.(2004)]{tafalla} Tafalla, M., Myers, P.C., Caselli, P.
and Walmsley, C.M.. 2004, A\& A , 416, 191.

\bibitem[Takahira et al.(2014)]{takahira} Takahira, K., Tasker, E.J. and Habe, A., 
2014, ApJ, 792, issue 1. or seen at arXiv:1407.4544. 

\bibitem[Testi et al.(2000)]{testi} Testi, L., Anneila, I. S.,
Luca, O. and Onello, J.S.,2000, ApJ , 540, pp. L53-L56.

\bibitem[Torii et al.(2011)]{torii} Torii, K., Enokiya, R., Sano, H., Yoshiike, S., 
Kanaoka, N., Ohama, A., Furukawa, N., Dawson, J.R., Moribe, N., Oishi, K., 
Nakashima, Y., Okuda, T., Yamamoto, H., Kawamura, A., Mizuno, N., Maezawa, H., 
Onishi, T., Mizuno, A. and Fukui, Y., 2011, ApJ, 738,46.

\bibitem[Truelove et al. (1997)]{truelove} Truelove, J.K., Klein, R.I.,
McKee, C.F., Holliman, J.H., Howell, L.H., and Greenough, J.A., 1997,
ApJ, 489, L179.

\bibitem[Vazquez-Semanedi et al. (2007)]{evaz} Vazquez-Semanedi, E., 
Gomez, G.C., Jappsen,A.K.,
Ballesteros-Paredes, J., Gonzalez, R. and Klessen, R.S., 2007, 
ApJ, 657, pp. 870-883.


\bibitem[Vishniac (1983)]{vishniac1983} Vishniac, E.T. 1983, ApJ, 274, 152.

\bibitem[Vishniac (1994)]{vishniac1994} Vishniac, E.T. 1994, ApJ, 428, 186.

\bibitem[Wang et al. (2004)]{wang} Wang, J.J., Chen, W.P., Miller, M., Qin, 
S.L., and Wu, Y.F.
2004, ApJ, 614, L105.

\bibitem[Whitehouse \& Bate (2006)]{whitehouse} Whitehouse,
S.C. and Bate, M.R. 2006, MNRAS, 367, 32

\bibitem[Whithworth \& Ward-Thompson (2001)]{whithworth} Whithworth,
A.P. and Ward-Thompson, D. 2001, ApJ, 547, 317

\bibitem[Whithworth \& Pongracic (1991)]{pongracic} Whithworth, A.P.,
and Pongracic, H. 1991,in {\it Fragmentation of molecular clouds and star
formation}, Eds. E. Falgarone, F. Boulangeer and G. Duvert,
International Astronomical Union, Kluwer, pp.  523.

\bibitem[Zinnecker \& Yorke (2007)]{zinne} Zinnecker, H. and Yorke, H.W.,  
(2007) , Ann. Rev.
A\& A, 45, 481.

\end{thebibliography}
\end{document}